\documentstyle[12pt,epsf]{article}
\newcommand{\del}{\partial}
\newcommand{\abs}[1]{{\left|{#1}\right|}}

\newcommand{\half}{{1 \over 2}}
\newcommand{\st}{{\tilde t}}
\newcommand{\sq}{{\tilde q}}
\newcommand{\absv}[1]{\left|#1\right|}
\newcommand{\Dmt}{{m_{\tilde q}^2}}
\newcommand{\cc}{{{3g_2^2-g_1^2}\over{12}}}
\newcommand{\gtsim}{\mathrel{\hbox{\raise0.2ex
\hbox{$>$}\kern-0.75em\raise-0.9ex\hbox{$\sim$}}}}
\newcommand{\ltsim}{\mathrel{\hbox{\raise0.2ex
\hbox{$<$}\kern-0.75em\raise-0.9ex\hbox{$\sim$}}}}
\newcommand{\lw}[1]{\smash{\lower2.0ex\hbox{#1}}}
\newcommand{\PRD}[3]{Phys. Rev. {\bf D{#1}} (19{#2}) {#3}}
\newcommand{\PRLet}[3]{Phys. Rev. Lett. {\bf {#1}} (19{#2}) {#3}}
\newcommand{\NPB}[3]{Nucl. Phys. {\bf B{#1}} (19{#2}) {#3}}
\newcommand{\PLB}[3]{Phys. Lett. {\bf B{#1}} (19{#2}) {#3}}
\newcommand{\PrTP}[3]{Prog. Theor. Phys. {\bf {#1}} (19{#2}) {#3}}
\makeatletter

\@addtoreset{equation}{section}
\makeatother
\begin{document}
\begin{titlepage}
\begin{flushright}
SAGA--HE--132\\
February 9,~1998
\end{flushright}
\vspace{50pt}
\centerline{\Large{\bf Spontaneous $CP$ Violation at Finite Temperature
in the MSSM}}
\vskip2.0cm
\begin{center}
{\bf Koichi~Funakubo$^{a,}$\footnote{e-mail: funakubo@cc.saga-u.ac.jp},
 Akira~Kakuto$^{b,}$\footnote{e-mail: kakuto@fuk.kindai.ac.jp},\\
 Shoichiro~Otsuki$^{b,}$\footnote{e-mail: otks1scp@mbox.nc.kyushu-u.ac.jp}
 and Fumihiko~Toyoda$^{b,}$\footnote{e-mail: ftoyoda@fuk.kindai.ac.jp}}
\end{center}
\vskip 1.0 cm
\begin{center}
{\it $^{a)}$Department of Physics, Saga University,
Saga 8408502 Japan}
\vskip 0.2 cm
{\it $^{b)}$Department of Liberal Arts, Kinki University in Kyushu,
Iizuka 8208255 Japan}
\end{center}
\vskip 1.5 cm
\centerline{\bf Abstract}
\vskip 0.2 cm
\baselineskip=15pt
By studying the effective potential of the MSSM at finite temperature,
we find that $CP$ can be spontaneously broken in the intermediate 
region between the symmetric and broken phases separated by the bubble
wall created at the phase transition. This type of $CP$ violation 
is necessary to have a bubble wall profile connecting $CP$ conserving 
vacua, while violating $CP$ halfway and generating sufficient baryon 
number without contradiction to the experimantal bounds on $CP$ violations.
Several conditions on the parameters in the MSSM are found for $CP$
to be broken in this manner.
\end{titlepage}
\baselineskip=18pt
\setcounter{page}{2}
\setcounter{footnote}{0}
\section{Introduction}
The idea of the electroweak baryogenesis\cite{reviewEB} is attractive 
in that it could solve the problem of matter-antimatter asymmetry in 
the universe by the knowledge of accessible experiments on the earth.
In particular, the nature of $CP$ violation, which is one of the 
requirements to generate the baryon asymmetry of the universe (BAU)
starting from the baryon-symmetric universe, will be revealed in
the near-future experiments.
The viable mechanisms of the electroweak baryogenesis, however, require
some extension of the standard model with other sources of $CP$ 
violation than the phase in the Kobayashi-Maskawa matrix.\par
Among the extensions, the minimal supersymmetric standard model (MSSM)
may not only admit various $CP$ violations but also cause first-order
electroweak phase transition (EWPT) with the small soft-supersymmetry-breaking
parameter in the stop mass-squared matrix\cite{CQW,DGG}.\  
It is also pointed out that the chargino and stop may play important 
roles in transporting the hypercharge into the symmetric phase, where
it biases the sphaleron process to generate the BAU\cite{EWB-MSSM}.\  
The $CP$-violating phases in the mass matrices are essential in
these scenarios, while they are constrained by various observations
such as the neutron electric dipole moment (EDM).
Another source of $CP$ violation, which was originally considered
in the baryogenesis mechanism\cite{NKC},\   is that in the Higgs sector,
that is, the relative phase of the expectation values of the two 
Higgs doublets. The Higgs VEVs including the phases, which characterize
the expanding bubble wall created at the first-order EWPT, vary spatially.
This spatially varying $CP$ violation makes, through the 
Yukawa coupling, the quarks and leptons to carry the hypercharge into
the symmetric phase\cite{FKOTTa}.\  
This scenario will work even if the superpartners are so heavy 
that they are not excited in the hot plasma.
Of course, since this $CP$-violating phase also enters the mass matrices
of the charginos, neutralinos, squarks and sleptons, this might enhance
the generated baryon number when they are thermally excited to act as
the charge carriers.\par
In a previous paper\cite{FKOTTb}\  we attempted to determine the profile 
of the bubble wall by solving the equations of motion for the effective 
potential at the transition temperature ($>T_C\simeq100\mbox{GeV}$) in
the two-Higgs-doublet model.
For some set of parameters, we presented a solution such that $CP$-violating 
phase spontaneously generated becomes as large as $O(1)$ 
around the wall while it completely vanishes in the broken and 
symmetric phase limits. We shall refer to this mechanism as
`transitional $CP$ violation'.
This solution gives a significant hypercharge flux, by the quark or 
lepton transport\cite{FKOTTc}.\  
We also showed that a tiny explicit $CP$ violation, which is consistent with 
the present bound on the neutron EDM, does nonperturbatively resolve the
degeneracy between the $CP$-conjugate pair of the bubbles to leave a sufficient
BAU after the EWPT\cite{FKOTa}.\par
In this paper, we examine the possibility of the transitional $CP$ 
violation at finite temperature in the MSSM by calculating the 
effective parameters, which are defined as the derivatives of the 
effective potential.
Similar analyses were executed by use of the high-temperature 
expansion of the finite-temperature corrections, one of which concerned
the problem of the order of the EWPT\cite{EMQ}\  and another focused on
the spontaneous $CP$ violation in the broken phase\cite{Comelli}.\  
It should be noted that the high-temperature expansion is not always 
a good approximation, especially when the masses of the particles running
through the loops are larger than the transition temperature. 
We apply it only to the light stop loop, while the contributions from
the other particles are treated numerically.
Although a tiny explicit $CP$ violation is needed to have nonzero 
BAU\cite{FKOTa},\  
we shall concentrate on the possibility of spontaneous $CP$ violation.
In \S~2, we briefly review the mechanism of the transitional $CP$ 
violation. We derive the formulas for the effective parameters,
which include both zero- and finite-temperature corrections, in \S~3.
We show the numerical results and analyze the possibility for the spontaneous 
$CP$ violation in \S~4. Discussions are given in \S~5.
The calculations of the loop corrections and the relevant integral formulas
are summarized in Appendix.
\section{Transitional $CP$ violation}
Consider a model with two Higgs doublets whose VEVs are parameterized 
as
\begin{equation}
  \langle\Phi_i\rangle
 =\left(\begin{array}{c} 0 \\
                        {1\over{\sqrt2}}\rho_i e^{i\theta_i}
        \end{array}\right),\qquad(i=1,2)
\end{equation}
and $\theta\equiv\theta_1-\theta_2$.
We assume that the gauge-invariant effective potential near the 
transition temperature has the form of\footnote{All the parameters 
in $V_{\rm eff}$ should be regarded as the effective ones containing
both zero- and finite-temperature corrections.}
\begin{eqnarray}
 V_{\rm eff}(\rho_i,\theta_i)&=&
 \half m_1^2\rho_1^2+\half m_2^2\rho_2^2 - m_3^2\rho_1\rho_2\cos\theta
 +{{\lambda_1}\over8}\rho_1^4+{{\lambda_2}\over8}\rho_2^4   \nonumber  \\
&+&{{\lambda_3+\lambda_4}\over4}\rho_1^2\rho_2^2
 + {{\lambda_5}\over4}\rho_1^2\rho_2^2\cos2\theta
 - \half(\lambda_6\rho_1^2+\lambda_7\rho_2^2)\rho_1\rho_2\cos\theta
    \nonumber  \\
&-& [A\rho_1^3+\rho_1^2\rho_2(B_0+B_1\cos\theta+B_2\cos2\theta)  
    \nonumber  \\
& &\qquad
    + \rho_1\rho_2^2(C_0+C_1\cos\theta+C_2\cos2\theta) + D\rho_2^3 ],
    \nonumber  \\
&=&\left[{\lambda_5\over2}\rho_1^2\rho_2^2
         -2(B_2\rho_1^2\rho_2+C_2\rho_1\rho_2^2)\right]  \nonumber\\
& &\qquad\times
   \left[\cos\theta-{{2m_3^2+\lambda_6\rho_1^2+\lambda_7\rho_2^2+
  2(B_1\rho_1+C_1\rho_2)}\over{2\lambda_5\rho_1\rho_2-8(B_2\rho_1+C_2\rho_2)}}
   \right]^2     \nonumber  \\
& & +\theta\mbox{-independent terms}.  \label{eq:Veff-ansatz}
\end{eqnarray}
The $\rho^3$-terms are expected to be induced at finite temperature 
in a model whose EWPT is of first order.
Since we do not consider any explicit $CP$ violation, all the 
parameters are assumed to be real.
For a given $(\rho_1,\rho_2)$, the spontaneous $CP$ violation occurs if
\begin{eqnarray}
  F(\rho_1,\rho_2) &\equiv&
  \displaystyle{
  {\lambda_5\over2}\rho_1^2\rho_2^2-2(B_2\rho_1^2\rho_2+C_2\rho_1\rho_2^2)
  }>0,       \label{eq:def-F}\\
 -1 <G(\rho_1,\rho_2) &\equiv&
 \displaystyle{
 {{2m_3^2+\lambda_6\rho_1^2+\lambda_7\rho_2^2+2(B_1\rho_1+C_1\rho_2)}\over
  {2\lambda_5\rho_1\rho_2-8(B_2\rho_1+C_2\rho_2)}}
 } <1.       \label{eq:def-G}
\end{eqnarray}
In the MSSM at the tree level, $\lambda_5=\lambda_6=\lambda_7=0$ and
$A=B_k=C_k=D=0$ ($k=0,1,2$), so that no spontaneous $CP$ violation occurs.
At zero temperature ($A=B_k=C_k=D=0$), it is argued that $\lambda_{5,6,7}$
are induced radiatively and (\ref{eq:def-F}) is satisfied if the 
contributions from the chargino and neutralino are large 
enough. For (\ref{eq:def-G}) to be satisfied, $m_3^2$ should be as small as
$\lambda_6\rho_1^2+\lambda_7\rho_2^2$ so that the pseudoscalar becomes 
too light\cite{Maekawa},\par
At $T\simeq T_C$, the values of $(\rho_1,\rho_2)$ vary
from $(0,0)$ to $(v_C\sin\beta_C,v_C\cos\beta_C)$ between the symmetric and 
broken phase regions, where the subscript $C$ denotes the quantities 
at the transition temperature. Then the effective parameters in 
(\ref{eq:Veff-ansatz}) include the temperature corrections as 
well. Hence there arises large possibility to satisfy both 
(\ref{eq:def-F}) and (\ref{eq:def-G}) in the intermediate region at
the transition temperature, without accompanying too light scalar.
If this is the case, a local minimum or a valley of $V_{\rm eff}$ 
appears for intermediate $(\rho_1,\rho_2)$ with a nontrivial 
$\theta$. It should be noted that such a local minimum need not to be
the global minimum of the effective potential.
For such a $V_{\rm eff}$ with appropriate effective 
parameters, the equations of motion for the Higgs fields predict that  
some class of solutions exist, which have $\theta$ of $O(1)$ in the 
intermediate region even if it vanishes in the broken phase\cite{FKOTTa}.\par
In the following sections, we calculate the effective parameters in
(\ref{eq:Veff-ansatz}) to examine whether the conditions (\ref{eq:def-F})
and (\ref{eq:def-G}) are satisfied for some intermediate $(\rho_1,\rho_2)$.
\section{Effective parameters of the MSSM}
Since we are concerned with the possibility of 
the spontaneous $CP$ violation, all the parameters in the lagrangian
are assumed to be real.
The tree-level Higgs potential of the MSSM is
\begin{eqnarray}
 V_0 &=&
 m_1^2\varphi_d^\dagger\varphi_d + m_2^2\varphi_u^\dagger\varphi_u
 +(m_3^2\varphi_u\varphi_d + \mbox{h.c})     \nonumber\\
 & &
 +{{\lambda_1}\over2}(\varphi_d^\dagger\varphi_d)^2
 +{{\lambda_2}\over2}(\varphi_u^\dagger\varphi_u)^2
 +\lambda_3(\varphi_u^\dagger\varphi_u)(\varphi_d^\dagger\varphi_d)
 +\lambda_4(\varphi_u\varphi_d)(\varphi_u\varphi_d)^* \nonumber\\
 & &
 +\left[
  {{\lambda_5}\over2}(\varphi_u\varphi_d)^2 +
  (\lambda_6\varphi_d^\dagger\varphi_d+\lambda_7\varphi_u^\dagger\varphi_u)
  \varphi_u\varphi_d + \mbox{h.c}
  \right],
\end{eqnarray}
where
\begin{eqnarray}
 m_1^2 &=& \tilde{m}_d^2+\absv{\mu}^2,\qquad
 m_1^2  =  \tilde{m}_u^2+\absv{\mu}^2,\qquad
 m_3^2  =  \mu B,   \nonumber\\
 \lambda_1&=&\lambda_2={1\over4}(g_2^2+g_1^2),\qquad
 \lambda_3 = {1\over4}(g_2^2-g_1^2),\qquad
 \lambda_4 = -\half g_2^2,    \\
 \lambda_5 &=& \lambda_6 = \lambda_7 = 0,
\end{eqnarray}
Here $g_{2(1)}$ is the $SU(2)$($U(1)$) gauge coupling, $\mu$ is the
coefficient of the Higgs quadratic interaction in the superpotential.
The mass squared parameters $\tilde{m}_{u,d}^2$ and $\mu B$ come from the
soft-supersymmetry-breaking terms so that they are arbitrary at this
level. $m_3^2$ could be complex but its phase can be eliminated by the
redefinition of the fields when $\lambda_5=\lambda_6=\lambda_7=0$.
We adopt the convention in which this $m_3^2$ is real and positive.\par
Let us parameterize the VEVs of the Higgs doublets as 
\begin{equation}
  \varphi_d
 ={1\over{\sqrt2}}\pmatrix{\rho_1\cr 0}
 ={1\over{\sqrt2}}\pmatrix{ v_1\cr 0},\qquad
  \varphi_u
 ={1\over{\sqrt2}}\pmatrix{ 0\cr \rho_2 e^{i\theta}}
 ={1\over{\sqrt2}}\pmatrix{ 0\cr v_2+iv_3}.  \label{eq:VEV}
\end{equation}
The effective potential at the one-loop level is
\begin{equation}
 V_{\rm eff}=V_0+V_1(\rho_i,\theta)+{\bar V}_1(\rho_i,\theta;T),
\end{equation}
where $V_1(\rho_i,\theta)$ is the zero-temperature correction given by
\begin{equation}
 V_1(\rho_i,\theta) =
 \sum_j\,n_j {{m_j^4}\over{64\pi^2}}
 \left[\log\left({{m_j^2}\over{M_{\rm ren}^2}}\right)-{3\over2}
 \right],       \label{eq:V1-zero-T}
\end{equation}
and ${\bar V}_1(\rho_i,\theta;T)$ is the finite temperature 
correction;
\begin{equation}
 {\bar V}_1(\rho_i,\theta;T) =
 {T^4\over2\pi^2}\sum_j\, n_j\int_0^\infty dx\,x^2
 \log\left[ 1 -{\rm sgn}(n_j)
  \exp\left(-\sqrt{x^2+m_j^2/T^2}\right)\right].
                \label{eq:V1-finite-T}
\end{equation}
Here we used the $\overline{\rm DR}$-scheme to renormalize $V_{\rm eff}$
with the renormalization scale $M_{\rm ren}$.
$n_j$ counts the degrees of freedom of each species including its 
statistics, that is, $n_j>0$ ($n_j<0$) for bosons (fermions).
$m_j$, which is a function of the Higgs background $(\rho_i,\theta)$,
is the mass eigenvalue of each species.\par
At the one-loop level, $\left(m_3^2\right)_{\rm eff}$ receives 
corrections only from the Higgs bosons, squarks, sleptons, and 
charginos and neutralinos. $\lambda_{5,6,7}$, which are zero at
the tree level, are generated only through the loops of these particles.
Among them, we consider the contributions of charginos($\chi^{\pm}$), 
neutralinos($\chi^0$), stops($\st$) and Higgs($\phi^{\pm}$).
The effective parameters are defined as the derivatives of $V_{\rm eff}$
at the origin of the order-parameter space:
\begin{eqnarray}
 \left(m_3^2\right)_{\rm eff} &=&
 -\left.{{\del^2 V_{\rm eff}}\over{\del v_1\del v_2}}\right|_0 =
 m_3^2 + \Delta_{\chi}m_3^2+ \Delta_{\tilde t}m_3^2 + 
               \Delta_{\phi^\pm}m_3^2,    \label{eq:def-eff-m32}\\
 \lambda_5 &=&
 \half\left(
  \left.{{\del^4 V_{\rm eff}}\over{\del v_1^2\del v_2^2}}\right|_0 -
  \left.{{\del^4 V_{\rm eff}}\over{\del v_1^2\del v_3^2}}\right|_0
 \right) =
 \Delta_{\chi}\lambda_5 + \Delta_{\tilde t}\lambda_5 +
               \Delta_{\phi^\pm}\lambda_5, \label{eq:def-eff-l5}\\ 
 \lambda_6 &=&
 -{1\over3}
   \left.{{\del^4 V_{\rm eff}}\over{\del v_1^3\del v_2}}\right|_0 =
 \Delta_{\chi}\lambda_6 + \Delta_{\tilde t}\lambda_6 +
               \Delta_{\phi^\pm}\lambda_6, \label{eq:def-eff-l6}\\ 
 \lambda_7 &=&
 -{1\over3}
   \left.{{\del^4 V_{\rm eff}}\over{\del v_1\del v_2^3}}\right|_0 = 
 \Delta_{\chi}\lambda_7 + \Delta_{\tilde t}\lambda_7 +
               \Delta_{\phi^\pm}\lambda_7. \label{eq:def-eff-l7}
\end{eqnarray}
The explicit forms of the corrections in term of the Feynman integrals 
are summarized in Appendix. In the following, we present the formulas
for these corrections from each species.\par
\noindent
{\bf (i) chargino and neutralino}\\
The mass matrices of the charginos and neutralinos are given by
\begin{eqnarray}
 M_{\chi^\pm} &=&
 \pmatrix{ M_2 & -{{ig_2}\over{\sqrt2}}\rho_2 e^{-i\theta} \cr
          -{{ig_2}\over{\sqrt2}}\rho_1 & -\mu }, 
                      \label{eq:chargino-mass-matrix}\\
 M_{\chi^0} &=&
 \pmatrix{
   M_2 & 0 & -{i\over2}g_2 \rho_1 & {i\over2}g_2\rho_2 e^{-i\theta} \cr
   0 & M_1 & {i\over2}g_1 \rho_1 & -{i\over2}g_1\rho_2 e^{-i\theta} \cr
  -{i\over2}g_2 \rho_1 &  {i\over2}g_1 \rho_1 & 0 & \mu  \cr
   {i\over2}g_2\rho_2 e^{-i\theta} & 
  -{i\over2}g_1\rho_2 e^{-i\theta} & \mu & 0 },
                      \label{eq:neutralino-mass-matrix}
\end{eqnarray}
respectively.
As noted in Appendix, all the contributions from the neutralinos
are proportional to those from the charginos, when the gaugino
mass parameters satisfy $M_2=M_1$. We assume this for simplicity.
Then the corrections from the charginos and neutralinos are summarized
as follows:
\begin{eqnarray}
 \Delta_{\chi}m_3^2 &=&
 -2g_2^2\left(1+{1\over{\cos^2\theta_W}}\right)\mu M_2\, L(M_2,\mu) 
 +{{g_2^2}\over{\pi^2}}\left(1+{1\over{\cos^2\theta_W}}\right)\mu M_2
  f_2^{(+)}\!\left({{M_2}\over T},{\mu\over T}\right), \nonumber\\
 & &              \label{eq:m32-gaugino}\\
 \Delta_{\chi}\lambda_5 &=&
  {{g_2^4}\over{8\pi^2}}\left(1+{2\over{\cos^4\theta_W}}\right)
     K\!\left({{M_2^2}\over{\mu^2}}\right)
 -{{g_2^4}\over{\pi^2T^4}}\left(1+{2\over{\cos^4\theta_W}}\right)\mu^2 M_2^2
     f_4^{(+)}\!\left({{M_2}\over T},{\mu\over T}\right), \nonumber\\
 & &              \label{eq:l5-gaugino}\\
 \Delta_{\chi}\lambda_6 &=&
 -{{g_2^4}\over{8\pi^2}}\left(1+{2\over{\cos^4\theta_W}}\right)
  {\mu\over{M_2}}\left[ - H\!\left({{M_2^2}\over{\mu^2}}\right)
      + K\!\left({{M_2^2}\over{\mu^2}}\right) \right]   \nonumber\\
 & &\qquad +
 {{g_2^4}\over{\pi^2}}\left(1+{2\over{\cos^4\theta_W}}\right) \left[
  {{\mu M_2}\over{T^2}}
          f_3^{(+)}\!\left({{M_2}\over T},{\mu\over T}\right)
  +{{\mu^3 M_2}\over{T^4}}
           f_4^{(+)}\!\left({{M_2}\over T},{\mu\over T}\right)\right]
                                  \nonumber\\
 &=& \Delta_{\chi}\lambda_7,        \label{eq:l6-gaugino}
\end{eqnarray}
where $L(M_2,\mu)$, $K(\alpha)$, $H(\alpha)$ and $f_{2,3,4}^{(+)}(a,b)$
are defined in Appendix.\par
\noindent
{\bf (ii) charged Higgs bosons}\\
The mass-squared matrix of the charged Higgs bosons is
\begin{equation}
 M^2_{\phi^\pm}=
 \pmatrix{
  m_1^2+{{g_2^2-g_1^2}\over8}\rho_1^2+{{g_2^2+g_1^2}\over8}\rho_2^2 &
  m_3^2+{1\over4}g_2^2\rho_1\rho_2e^{i\theta}   \cr
  m_3^2+{1\over4}g_2^2\rho_1\rho_2e^{-i\theta}   &
  m_2^2+{{g_2^2+g_1^2}\over8}\rho_1^2+{{g_2^2-g_1^2}\over8}\rho_2^2}.
\end{equation}
The low-energy parameters in this matrix are arranged to
break the gauge symmetry, so that one of the mass-squared eigenvalues
evaluated at $\rho_i=0$ should be negative; $m_1^2m_2^2-m_3^4<0$.
This negative mass squared makes the finite-temperature corrections to
the effective potential complex for small $\rho_i$.
(Suppose negative $m_j^2$ in (\ref{eq:V1-finite-T}).) 
This pathology will be cured by taking the higher-order corrections
into account\cite{DJ}.\  
Among the corrections, the so-called `daisy diagrams' are the most
dominant ones at high temperature since they grow as $T^2$.
Hence we replace $m_1^2$ and $m_2^2$ in the Higgs loops with the 
`daisy-corrected' ones given by
\begin{equation}
  {\bar m}_1^2 = m_1^2 +k_1 T^2,\qquad
  {\bar m}_2^2 = m_2^2 +k_2 T^2,       \label{eq:def-m-bar}
\end{equation}
where
\begin{eqnarray}
 k_1 &=& {1\over{16\pi^2}}(3g_2^2+g_1^2),   \nonumber\\
 k_2 &=& {1\over{16\pi^2}}(3g_2^2+g_1^2+4y_t^2).
\end{eqnarray}
This determines the limiting temperature $T_{\rm low}$, under which
the origin of the effective potential is not a local minimum, that is, 
${\bar m}_1^2{\bar m}_2^2<(m_3^2)^2$:
\begin{equation}
 T_{\rm low}^2 =
 {{\sqrt{(k_1m_2^2-k_2m_1^2)^2+4k_1k_2(m_3^2)^2}-(k_1m_2^2+k_2m_1^2)}\over
  {2k_1k_2}}.      \label{eq:def-T-low}
\end{equation}
This limiting temperature is rather large for $\tan\beta_0>2$.
It will be shown numerically that the Higgs 
contributions to the effective parameters are smaller by order 
four or five than the others as long as they are well defined.
Hence we simply neglect the Higgs contributions 
when $T<T_{\rm low}$ in the following calculations.
Even if the approximation by simply substituting ${\bar m}_{1,2}^2$ 
for $m_{1,2}^2$ in the Higgs loops is not justified, 
the origin of $V_{\rm eff}$ should be a local minimum at $T\simeq T_C$
as long as the EWPT is of first order. Then the effective $m_1^2m_2^2$
would be large enough so that the contributions from the Higgs would become
small, as is clear from the following formulas.
\begin{eqnarray}
 \Delta_{\phi^\pm}m_3^2 &=&
 \half g_2^2 m_3^2\, L(\mu_1,\mu_2) +
 {1\over{4\pi^2}}g_2^2m_3^2 f_2^{(-)}\!\left({{\mu_1}\over T},{{\mu_2}\over T}\right),
                         \label{eq:m32-higgs}\\
 \Delta_{\phi^\pm}\lambda_5 &=&
 -{{g_2^4}\over{64\pi^2}}{{m_3^4}\over{\mu_1^2\mu_2^2}}
    K\!\left({{\mu_1^2}\over{\mu_2^2}}\right)
 -{1\over{8\pi^2T^4}}g_2^4 m_3^4
  f_4^{(-)}\!\left({{\mu_1}\over T},{{\mu_2}\over T}\right),
  \label{eq:l5-higgs}\\
 \Delta_{\phi^\pm}\lambda_6 &=&
 {{g_2^4}\over{64\pi^2}}{{m_3^2}\over{\mu_1^2}}\Biggl\{
  - H\!\left({{\mu_1^2}\over{\mu_2^2}}\right)
  +\Bigl[ 1 -{{m_1^2}\over{2\mu^2\cos^2\theta_W}}
          -\left(1-{1\over{2\cos^2\theta_W}}\right)
       {{m_2^2}\over{\mu_2^2}}\Bigr]
      K\!\left({{\mu_1^2}\over{\mu_2^2}}\right) \Biggr\}\nonumber\\
 & & +
 {{g_2^4m_3^2}\over{8\pi^2T^2}}\Biggl[
  f_3^{(-)}\!\left({{\mu_1}\over T},{{\mu_2}\over T}\right)\! +
   \Bigl( {{\mu_2^2}\over{T^2}}-{{m_1^2}\over{2T^2\cos^2\theta_W}}
                                      \nonumber\\
 & &\qquad\qquad\qquad
  -\left(1-{1\over{2\cos^2\theta_W}}\right)\!{{m_2^2}\over{T^2}}\Bigr)
       \! f_4^{(-)}\!\left({{\mu_1}\over T},{{\mu_2}\over T}\right)
             \Biggr],   \label{eq:l6-higgs}\\
 \Delta_{\phi^\pm}\lambda_7 &=&
 {{g_2^4}\over{64\pi^2}}{{m_3^2}\over{\mu_1^2}}\Biggl\{
 - H\!\left({{\mu_1^2}\over{\mu_2^2}}\right)
 +\Bigl[1-\left(1-{1\over{2\cos^2\theta_W}}\right){{m_1^2}\over{\mu_2^2}}
 -{{m_2^2}\over{2\mu_2^2\cos^2\theta_W}} \Bigr]  
       K\!\left({{\mu_1^2}\over{\mu_2^2}}\right) \Biggr\}\nonumber\\
 & &+
 {{g_2^4m_3^2}\over{8\pi^2T^2}}\Biggl[
  f_3^{(-)}\!\left({{\mu_1}\over T},{{\mu_2}\over T}\right) +
  \Bigl( {{\mu_2^2}\over{T^2}}
    -\left(1-{1\over{2\cos^2\theta_W}}\right)\!{{m_1^2}\over{T^2}}
    \nonumber\\
 & &\qquad\qquad\qquad
  -{{m_2^2}\over{2T^2\cos^2\theta_W}} \Bigr)\!
    f_4^{(-)}\!\left({{\mu_1}\over T},{{\mu_2}\over T}\right) 
             \Biggr],   \label{eq:l7-higgs}
\end{eqnarray}
where
\begin{equation}
 \mu_{1,2}^2 = 
 {{{\bar m}_1^2+{\bar m}_2^2\pm\sqrt{({\bar m}_1^2-{\bar m}_2^2)^2+4m_3^4}}\over2}.
\end{equation}
The definitions of the various functions used in the above formulas
are given in Appendix.\par
\noindent
{\bf (iii) light stop and $\rho^3$-term}\\
When $m_j^2$ in (\ref{eq:V1-finite-T}) vanishes as
$\rho_i\rightarrow0$, the second and higher derivatives of it for
the bosonic loops are ill-defined at $\rho_i=0$.
This singularity originates from the zero mode in the summation over
the Matsubara modes.
Upon approximated by the high-temperature expansion\cite{DJ},\  
(\ref{eq:V1-finite-T}) receives $\rho^3$-terms with positive 
coefficients from the bosonic particles whose masses behave as 
$O(\rho^2)$ for $\rho_i\simeq 0$.
This $\rho^3$-terms are supposed to make the EWPT first order.
In the MSSM, the candidates generating such terms are the weak gauge bosons,
the Higgs bosons and the scalar partner of the quarks and leptons with
appropriate mass parameters.
Among them the Higgs bosons and the squarks and sleptons could yield
$\theta$-dependent $\rho^3$-terms, that is, $B_{1,2}$ and/or $C_{1,2}$ 
in (\ref{eq:Veff-ansatz}), which will affect the conditions (\ref{eq:def-F})
and (\ref{eq:def-G}), if their mass eigenvalues vanishes as
$\rho_i\rightarrow0$.
When one of the soft-supersymmetry-breaking mass parameters in the 
squark mass matrix vanishes, this situation is realized.
Here we consider only the top squark (stop) because of its large 
Yukawa coupling.\par
The mass-squared matrix of the stop is
\begin{equation}
 M_\st^2 = \pmatrix{ m_{11}^2 & m_{12}^2 \cr
                     m_{12}^{2*} & m_{22}^2 \cr}, \label{eq:def-Mt}
\end{equation}
where
\begin{eqnarray}
 m_{11}^2&=&
  m_{\tilde q}^2-{1\over8}\left({{g_1^2}\over3}-g_2^2\right)
  (\rho_1^2-\rho_2^2)+\half y_t^2\rho_2^2,  \label{eq:def-m11}\\
 m_{22}^2&=&
  m_\st^2+{1\over6}g_1^2(\rho_1^2-\rho_2^2)+\half y_t^2\rho_2^2,
                                                  \label{eq:def-m22}\\
 m_{12}^2&=&
  {{y_t}\over{\sqrt{2}}}\left[(\mu \rho_1+A_t\rho_2\cos\theta)-
 iA_t\rho_2\sin\theta\right].
                                                  \label{eq:def-m12}
\end{eqnarray}
Here $m_{\tilde q}^2$, $m_\st^2$ and $A_t$ come from the 
soft-supersymmetry-breaking terms and $y_t$ is the top Yukawa coupling.
Although the relative phase between $\mu$ and $A_t$ yields an explicit
$CP$ violation, we assume they are real.\par
The temperature-dependent part of the stop contribution to the effective
potential is
\begin{equation}
 {\bar V}_\st(\rho_i,\theta;T) =
 3{{T^4}\over{2\pi^2}}\left[2I_B(a_+^2) + 2I_B(a_-^2)\right],
                                                 \label{eq:stop-Veff}
\end{equation}
where $I_B(a^2)$ is defined by
\begin{equation}
  I_B(a^2) =
  \int_0^\infty dx\, x^2\log\left(1-e^{-\sqrt{x^2+a^2}}\right),     
            \label{eq:IB}
\end{equation}
and the factor 2 counts the degrees of freedom of complex scalars
and $a_\pm^2\equiv m_\pm^2/T^2$ with
\begin{equation}
 m_\pm^2\equiv\half\left[m_{11}^2+m_{22}^2\pm
              \sqrt{(m_{11}^2-m_{22}^2)^2+4\abs{m_{12}^2}^2}\right]
          \label{eq:def-m-pm}
\end{equation}
being the eigenvalues of $M_\st^2$.
Since $\mu$ and $A_t$ are real, $\mu A_t\not=0$ is required to give 
$\theta$-dependence in $m_\pm^2$.
If $m_\sq$ and/or $m_\st$ vanishes, $m_+^2$ and/or $m_-^2$ behave as
$O(\rho^2)$ for $\rho\simeq0$. We assume $m_\st\simeq0$ and
$m_{\sq}\gg T$, since too light left-handed stop, which couples to the
$SU(2)$ gauge bosons, would lead to a large correction to the $\rho$-parameter.
Then $m_{\pm}$ is given approximately as
\begin{eqnarray}
 m_+^2 &\simeq&
 m_{\sq}^2+{1\over8}\left(-{{g_1^2}\over3}+g_2^2\right)(\rho_1^2-\rho_2^2)+
 \half y_t^2\rho_2^2 + 
 \left({Y\over{4m_{\sq}^2}}-{XY\over{4m_{\sq}^4}}-{Y^2\over{8m_{\sq}^6}}
 \right)     \nonumber\\
 &\equiv& m_{\sq}^2+m_{\rho_+}^2,   \label{eq:def-m-}\\
 m_-^2 &\simeq&
 {1\over6}g_1^2(\rho_1^2-\rho_2^2)+\half y_t^2\rho_2^2-
 \left({Y\over{4m_{\sq}^2}}-{XY\over{4m_{\sq}^4}}-{Y^2\over{8m_{\sq}^6}}
 \right)         \nonumber\\
 &\equiv&m_{\rho_-}^2,              \label{eq:def-m+}
\end{eqnarray}
where
\begin{eqnarray}
 X &=&
 {1\over8}\left(-{5\over3}g_1^2+g_2^2\right)(\rho_1^2-\rho_2^2),
                         \label{eq:def-P2}\\
 Y&=&
 2y_t^2(\mu^2\rho_1^2+2\mu A_t\rho_1\rho_2\cos\theta+ A_t^2\rho_2^2).
                         \label{eq:def-Q2}
\end{eqnarray}
To evaluate the stop contributions to the effective parameters defined
in (\ref{eq:def-eff-m32}) -- (\ref{eq:def-eff-l7}), we need the behavior
of $V_\st(\rho_i,\theta;T)$ at $\rho_i\simeq0$.
For this purpose, we employ the Taylor expansion around
$a_\sq^2=m_\sq^2/T^2$ to evaluate $I_B(a_+^2)$:
\begin{equation}
 I_B(a_+^2) =
 I_B(a_{\sq}^2)+I_B'(a_{\sq}^2)a_{\rho_+}^2+
 {I_B''(a_{\sq}^2)\over2}(a_{\rho_+}^2)^2+\ldots,
\end{equation}
where $a_{\rho_+}^2=m_{\rho_+}^2/T^2$. On the other hand, we use
the high-temperature expansion for $I_B(a_-^2)$:
\begin{equation}
 I_B(a_-^2) =
 -{\pi^4\over45}+{\pi^2\over12}a_{\rho_-}^2 -{\pi\over6}a_{\rho_-}^3 +
 \lambda_- a_{\rho_-}^4+\ldots,
\end{equation}
where up to $O(a^3)$ the coefficients are obtained from the well-known
formula\cite{DJ},\  which includes $a^4\log a$-term in addition to 
$a^4$-term. In exchange for dropping the $a^4\log a$-term, we have
decided $\lambda_-=0.1764974$ by numerical fitting for $0<a_-<1$.\par
We obtain the stop corrections to the effective parameters:
\begin{eqnarray}
 \Delta_{\st}m_3^2 &=&
  N_c y_t^2\mu A_t\, L(m_\sq,0) +
  {{3T^2}\over{\pi^2}}{{y_t^2\mu A_t}\over\Dmt}\left[
   I_B^\prime(a_\sq^2)-{{\pi^2}\over12}\right],     \label{eq:m32-stop2}\\
 \Delta_{\st}\lambda_5 &=&
 -{{N_cy_t^4}\over{16\pi^2}}{{\mu^2A_t^2}\over{m_\sq^2M_{IR}^2}}
     K\left({{m_\sq^2}\over{M_{IR}^2}}\right)  \nonumber\\
 & &{}+
 {{N_cy_t^4\mu^2A_t^2}\over{\pi^2(\Dmt)^2}}\left[
  {{2T^2}\over\Dmt}\left( 
     -I_B'(a_\sq^2)+{\pi^2\over12} \right) + I_B''(a_\sq^2) + 2\lambda_- 
     \right],                 \label{eq:l5-stop2}\\
 \Delta_{\st}\lambda_6 &=&
  {{N_cy_t^2}\over{16\pi^2}}{{\mu A_t}\over{m_\sq^2}}\left[
  {1\over4}\left({{g_1^2}\over3}-g_2^2\right) -
  {{g_1^2m_\sq^2}\over{3M_{IR}^2}}H\left({{M_{IR}^2}\over{m_\sq^2}}\right)+
  {{y_t^2\mu^2}\over{M_{IR}^2}}
     K\left({{m_\sq^2}\over{M_{IR}^2}}\right) \right] \nonumber\\
 & &+
 {{N_cy_t^2\mu A_t}\over{\pi^2\Dmt}}\left\{
   {{2T^2}\over\Dmt}\left({{y_t^2\mu^2}\over\Dmt}+(-{5\over3}g_1^2+g_2^2)
      \right)\left[I_B'(a_\sq^2)-{\pi^2\over12}\right]\right.\nonumber\\
 & &\qquad
 \left. -\left({{y_t^2\mu^2}\over\Dmt}+\cc\right)I_B''(a_\sq^2)
     +2\left({{g_1^2}\over3}-{{y_t^2\mu^2}\over\Dmt}
       \right)\lambda_- \right\},          \label{eq:l6-stop2}\\
 \Delta_{\st}\lambda_7 &=&
  {{N_cy_t^2}\over{16\pi^2}}{{\mu A_t}\over{m_\sq^2}}\Biggl[
  -\left(y_t^2+{1\over4}\left({{g_1^2}\over3}-g_2^2\right)\right)
   \nonumber\\
  & &{}-\left(y_t^2-{{g_1^2}\over3}\right){{m_\sq^2}\over{M_{IR}^2}}
        H\left({{M_{IR}^2}\over{m_\sq^2}}\right)
  +{{y_t^2 A_t^2}\over{M_{IR}^2}}
      K\left({{m_\sq^2}\over{M_{IR}^2}}\right) \Biggr] \nonumber\\
  & &{}+
 {{N_cy_t^2\mu A_t}\over{\pi^2\Dmt}}\Biggl\{
   {{2T^2}\over\Dmt}\left({{y_t^2\mu^2}\over\Dmt}-(-{5\over3}g_1^2+g_2^2)
 \right)\!\!
     \left[I_B'(a_\sq^2)-{\pi^2\over12}\right] \nonumber\\
 & &\qquad
  -\left(y_t^2+{{y_t^2\mu^2}\over\Dmt}-\cc\right)\!
        I_B''(a_\sq^2)
   +2\left(y_t^2-{{g_1^2}\over3}-{{y_t^2\mu^2}\over\Dmt}\right)\lambda_-
     \Biggr\},          \label{eq:l7-stop2}
\end{eqnarray}
where $N_c=3$ and $M_{IR}$ is the infrared cutoff parameter, 
which will be taken to be the order of the transition temperature
This is needed because of the infrared singularity encountered in
the presence of a massless particle through the loops, as is well 
known\cite{CW}.\  
This is cured by calculating the fourth derivatives away from the 
origin. Then, by minimizing the effective potential, the mass scale in
the logarithm is replaced by the VEV, that is, the dimensional transmutation 
occurs. We have checked that as long as $M_{IR}\gtsim100\mbox{GeV}$, 
$M_{IR}$-dependence is not so significant that we simply use $M_{IR}$ 
instead of minimizing the $V_{\rm eff}$.\par
Now we are ready to extract $\rho^3$-term in the stop contribution, 
which is given by. 
\begin{equation}
 \left.{\bar V}_\st(\rho_i,\theta;T)\right|_{O(\rho^3)} =
 3{{T^4}\over{\pi^2}}\,\left.I_B(a_-^2)\right|_{O(a^3)} \simeq
 -{{T^4}\over{2\pi}}\left(a_{\rho_-}^2\right)^{3/2}.  \label{eq:Vst-a3}
\end{equation}
When
$A_t/m_{\tilde q}, \mu/m_{\tilde q} ,(g_1/y_t)^2\ll 1$,
we pick up higher-order terms to obtain
\begin{eqnarray}
 \left.{\bar V}_\st(\rho_i,\theta;T)\right|_{O(\rho^3)}
 &\simeq&
 -{T\over{4\sqrt{2}\pi}}\abs{y_t}^3\left\{
 \left[1-{{g_1^2}\over{2y_t^2}}
       -{3\over2}\left({A_t\over{m_{\sq}}}\right)^2
       +{3\over8}\left({{A_t}\over{m_{\sq}}}\right)^4
 \right]\rho_2^3                     
 \right. \nonumber\\
 & &\quad
 +\left({{g_1^2}\over{2y_t^2}}-{{3\mu^2}\over{2m_{\sq}^2}}\right)
  \rho_1^2\rho_2 
 +{3\over2}\left({{\mu A_t}\over{m_{\tilde q}^2}}\right)^2
   \rho_1^2\rho_2\cos^2\theta           \nonumber\\
 & &\quad\left.
 +{{3\mu A_t}\over{m_{\tilde q}^2}}\left[
   -1+\half \left({A_t\over{m_{\sq}}}\right)^2+{1\over6}
 \left(g_1\over{y_t}\right)^2\right]
        \rho_1\rho_2^2\cos\theta   + \cdots \right\}.
       \label{eq:Vst-v3-1-exp2}
\end{eqnarray}
From this expansion, we extract 
\begin{eqnarray}
 C_1 &\simeq&
 {T\over{4\sqrt2\pi}}{\absv{y_t}}^3{{3\mu A_t}\over{m_{\tilde q}^2}}
 \left[-1+{1\over2}\left({A_t \over m_{\tilde q}}\right)^2 +
 {1\over6}\left({g_1 \over y_t}\right)^2\right],    \label{eq:C1-stop}\\
 B_2 &\simeq&
 {T\over {4\sqrt2\pi}}{\absv{y_t}}^3{3\over2}
 \left({{\mu A_t}\over{m_{\tilde q}}^2}\right)^2.   \label{eq:B2-stop}
\end{eqnarray}
Because $\absv{\mu A_t/m_\sq^2}\ll 1$, it is expected that $B_2$
is much smaller than $C_1$.
We can neglect $B_1$ and $C_2$, which will be induced from the sbottom 
loops, compared with $C_1$ and $B_2$, because the bottom Yukawa 
coupling $\absv{y_b}$ is much smaller than $\absv{y_t}$.\par
\section{Numerical Results}
We examine whether the conditions $G(\rho_1,\rho_2)<1$ and 
$\absv{G(\rho_1,\rho_2)}<1$ are satisfied or not by evaluating the
effective parameters included in $F$ and $G$. 
As note above, we numerically calculates the integrals
in the finite-temperature corrections. 
Before showing the numerical results, we comment on
some general properties of the behavior of the parameters.\par
If only the light stop contributes to the $\theta$-dependent $\rho^3$-terms,
$C_2=0$ so that $F(\rho_1,\rho_2)>0$ holds for $\rho_2$ satisfying
\begin{equation}
  (\lambda_5 \rho_2 - 4B_2)\rho_1 > 0.
\end{equation}
As long as we take $\rho_1$ to be positive, this implies
\begin{eqnarray}
 & &\rho_2 > {{4B_2}\over{\lambda_5}},\qquad\mbox{for }\lambda_5>0 \\
 & &{{4B_2}\over{\lambda_5}}<\rho_2<0,\qquad\mbox{for }\lambda_5<0.
\end{eqnarray}
The latter case corresponds to a negative $\tan\beta$ at finite 
temperature. Since we adopt $\tan\beta_0>0$ at the tree level in
the following examples and we found several $CP$-violating bubble wall 
solutions for $\lambda_5>0$\cite{FKOTTa},\  we concentrate on the 
former case here.
Note that positive corrections to $\lambda_5$ come from the charginos
and neutralinos at zero temperature and the light stop at finite 
temperature. All the other contributions are always negative.
For the expansion (\ref{eq:Vst-v3-1-exp2}) to be valid, 
$\absv{\mu A_t/m_\sq^2}\ll 1$ so that we expect that the stop 
contribution is much smaller than those from the charginos and 
neutralinos. The maximum is realized around $\mu^2=M_2^2$, which 
corresponds to the maximum of $K(M_2^2/\mu^2)$.
Since $K(\alpha)$ slowly varies around the peak, 
$\Delta_\chi\lambda_5(T=0)$ is positive for a rather wide range of 
$M_2^2/\mu^2$. We have checked that even if the finite-temperature 
corrections are taken into account, $\lambda_5$ is positive for
$1/2\ltsim M_2^2/\mu^2\ltsim 3$ for $T\ltsim 100\mbox{GeV}$.
In fact, $\lambda_5\gtsim 10^{-4}$ at $T\simeq100\mbox{GeV}$ and
$B_2=O(10^{-5})T\simeq 10^{-3}\mbox{GeV}$ in our examples, so that 
the condition $F(\rho_1,\rho_2)>0$ is satisfied for 
$\rho_2\gtsim 10\mbox{GeV}$. This will not impose a strong constraint as 
long as $\tan\beta_C$ is not so small. On the other hand,
for $\absv{G(\rho_1,\rho_2)}<1$ to be satisfied, the value of the
tree-level $m_3^2$ must be tuned since 
the magnitude of $\left(m_3^2\right)_{\rm eff}$ must be the same order
as $\lambda_6\rho_1^2+\lambda_7\rho_2^2$ and $C_1\rho_2$, 
which are radiatively induced at finite temperature.\par
Now we examine the condition $\absv{G(\rho_1,\rho_2)}<1$ for several 
sets of the tree-level parameters, which are $m_1^2$, $m_2^2$, $m_3^2$,
$A_t$, $M_2$, $\mu$, $m_\sq$ and $m_\st=0$, at temperatures of
$O(100\mbox{GeV})$.
Instead of giving $m_1^2$ and $m_2^2$, we input the values of 
the tree-level $\tan\beta_0$ and the absolute value of the Higgs VEV
$v_0$, which are related to the masses-squared parameters by
the relations defining the minimum of $V_0$:
\begin{eqnarray}
 m_1^2 &=& m_3^2\tan\beta_0-{1\over2}m_Z^2\cos(2\beta_0), \nonumber\\
 m_2^2 &=& m_3^2\cot\beta_0+{1\over2}m_Z^2\cos(2\beta_0).
\end{eqnarray}
As seen from the formulas for the effective parameters,
the signs of the corrections depend on those of $\mu M_2$ and $\mu A_t$.
For example, the sign of $\Delta_\chi m_3^2(T=0)$ is the same as that of
$\mu M_2$, while the temperature corrections to it have the opposite sign.
If we take $\absv{A_t}\ll\absv{M_2}\sim m_\sq$, the chargino and 
neutralino contributions dominate over those from the stops.
As long as we adopt a positive $m_3^2$ at the tree-level, negative
$\mu M_2$ is needed to have a nearly zero 
$\left(m_3^2\right)_{\rm eff}$ at $T\simeq T_C$.
Since $f_2^{(\mp)}(m_1/T,m_2/T)$ is a positive and increasing 
function of $T$, the finite-temperature part of $\Delta_\chi m_3^2(T)$
works to reduce $\left(m_3^2\right)_{\rm eff}$ to almost zero from
its positive zero-temperature value.\par
Hence we take two parameter sets with $\mu A_t>0$ and $\mu A_t<0$,
respectively. For each case, the $T$-dependences of the effective 
parameters are studied and $\absv{G(\rho_1,\rho_2)}$ is plotted in
$(\rho_1,\rho_2)$-plane at several temperatures with $\tan\beta_0=1.2$
and $\tan\beta_0=5$.
All the numerical values having mass dimension 
should be understood to be in the unit of GeV.
We use $v_0=246$, $m_t=177$ and $M_{\rm ren}=M_{IR}=100$ in
these examples.\par
\noindent
{\bf (i) $\mu A_t>0$}\\
The parameters in the first example is given in Table~\ref{table:1}.
\par
\begin{figure}
 \epsfxsize=11.6cm
 \centerline{\epsfbox{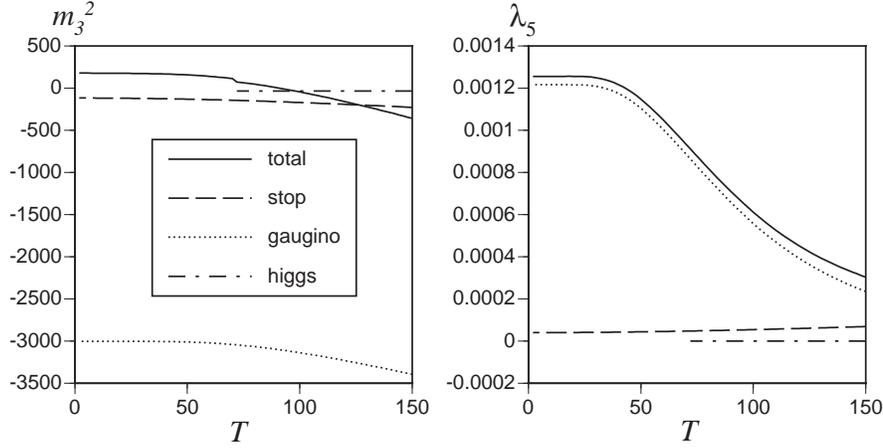}}
 \caption{$\left(m_3^2\right)_{\rm eff}$ and $\lambda_5$ as functions
 of temperature $T$. The total values are given by the solid curves, the 
 corrections from the stop, chargino-neutralino and the charged Higgs 
 bosons are depicted by the dashed, dotted and dotted-dashed curves, 
 respectively. For $T<T_{\rm low}=71.4$ the Higgs contributions are
 ignored.
 }
 \label{fig:1}
\end{figure}

\begin{figure}
 \epsfxsize=11.6cm
 \centerline{\epsfbox{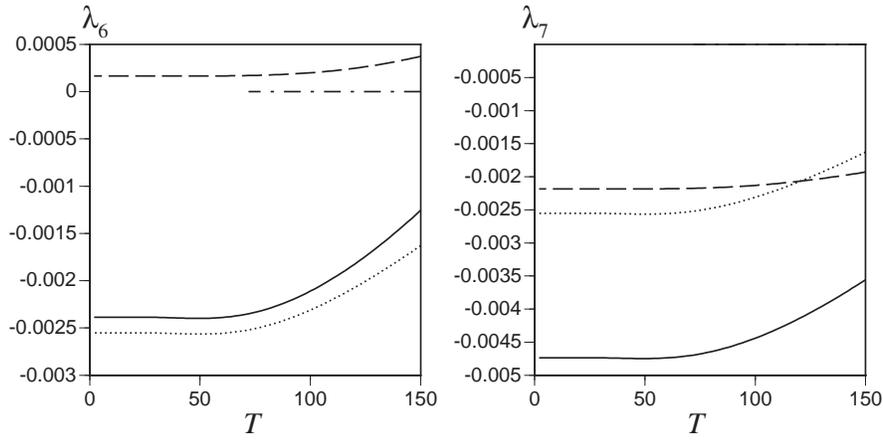}}
 \caption{$\lambda_6$ and $\lambda_7$ as functions
 of temperature.
 }
 \label{fig:2}
\end{figure}

\begin{figure}
 \epsfxsize=11.6cm
 \centerline{\epsfbox{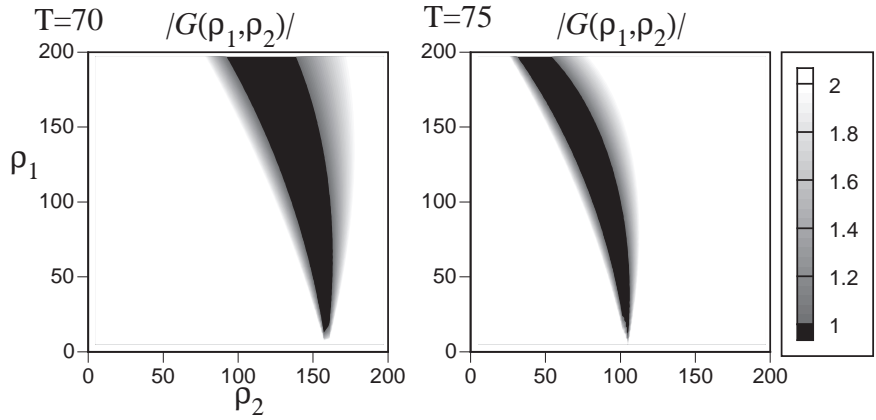}}
 \caption{Contour plots of $\absv{G(\rho_1,\rho_2)}$ at $T=70$ and 
 $75$. $\absv{G(\rho_1,\rho_2)}$ is satisfied in the black region.
 }
 \label{fig:3}
\end{figure}

\begin{figure}
 \epsfxsize=11.6cm
 \centerline{\epsfbox{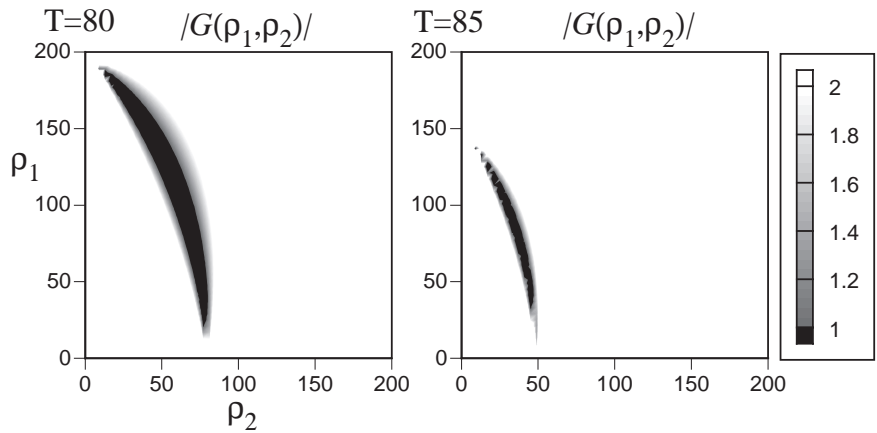}}
 \caption{The same as Fig.~\protect\ref{fig:3} at $T=80$ and $85$. 
 }
 \label{fig:4}
\end{figure}

\noindent
\begin{table}{l}{h}
 \caption{The parameters used in the numerical analysis in the case
 of $\mu A_t>0$.}
 \label{table:1}
 \begin{center}
  \begin{tabular}{ccccc}
   \hline
    $m_3^2$&$A_t$& $M_2$ & $\mu$ &$m_\sq$ \\
   \hline
    $3300$ & $10$& $-400$ &$200$ &  $400$ \\
   \hline
  \end{tabular}
 \end{center}
\end{table}
(1) $\tan\beta_0=1.2$\\
For these parameters, $T_{\rm low}=71.411$, below which we neglect
the contributions of the charged Higgs bosons.
As shown in Figs.~\ref{fig:1} and \ref{fig:2}, their values are much smaller 
compared to the others so that we expect them not to alter the results
significantly.
In this case, we have
\begin{equation}
 B_2/T = 1.532346\times10^{-5},\qquad
 C_1/T = -4.846485\times10^{-3}.
\end{equation}
As seen from the curves in Fig.~\ref{fig:2}, $\lambda_{6,7}=O(10^{-3})$
so that $C_1\rho_2$ is comparable to $\lambda_6\rho_1^2+\lambda_7\rho_2^2$
for $\rho_i\sim100$ at $T\simeq100$. Hence, when 
$\absv{G(\rho_1,\rho_2)}<1$, $\left(m_3^2\right)_{\rm eff}$ can be
larger compared to the case of spontaneous $CP$ violation at $T=0$
if $C_1$ has the same sign as it, which is the present case.
From (\ref{eq:C1-stop}), $C_1$ has the opposite sign to $\mu A_t$.
At $T=100$, $\lambda_5=6.11026351\times10^{-4}$ so that 
$F(\rho_1,\rho_2)>0$ for $\rho_2>4B_2/\lambda_5=10.03$.
$\absv{G(\rho_1,\rho_2)}$ is plotted for $T=70$, $75$, $80$ and $85$.
There exists a region where the condition $\absv{G(\rho_1,\rho_2)}<1$
is satisfied, as shown in Figs.~\ref{fig:3} and \ref{fig:4} at each
temperature.\par
\begin{figure}
 \epsfxsize=11.6cm
 \centerline{\epsfbox{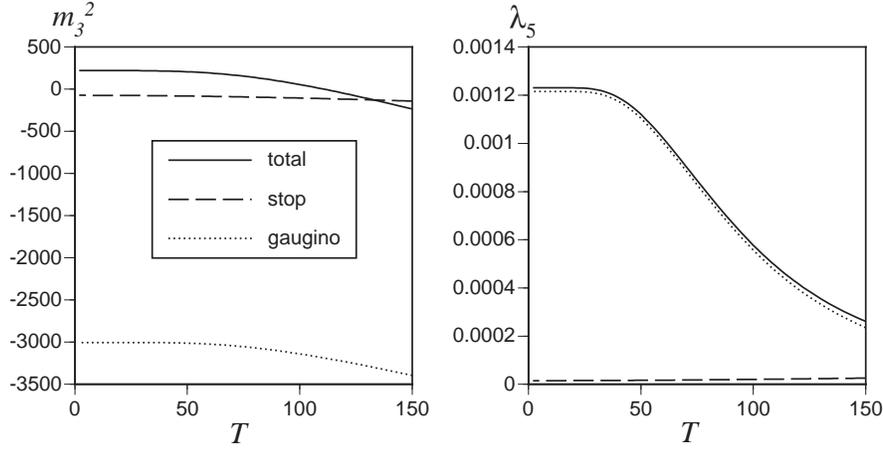}}
 \caption{$\left(m_3^2\right)_{\rm eff}$ and $\lambda_5$ as functions
 of temperature. The total values are given by the solid curves, the 
 corrections from the stop and chargino-neutralino are depicted by
 the dashed and dotted curves, respectively.
 }
 \label{fig:5}
\end{figure}

\begin{figure}
 \epsfxsize=11.6cm
 \centerline{\epsfbox{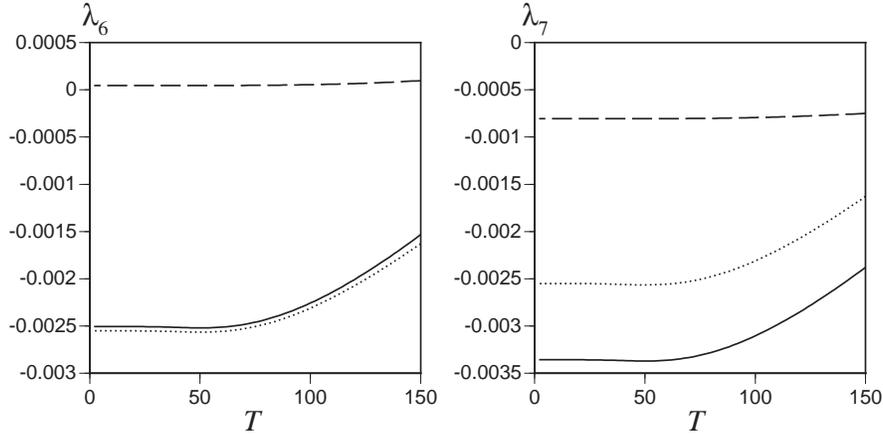}}
 \caption{$\lambda_6$ and $\lambda_7$ as functions
 of temperature.
 }
 \label{fig:6}
\end{figure}

\begin{figure}
 \epsfxsize=11.6cm
 \centerline{\epsfbox{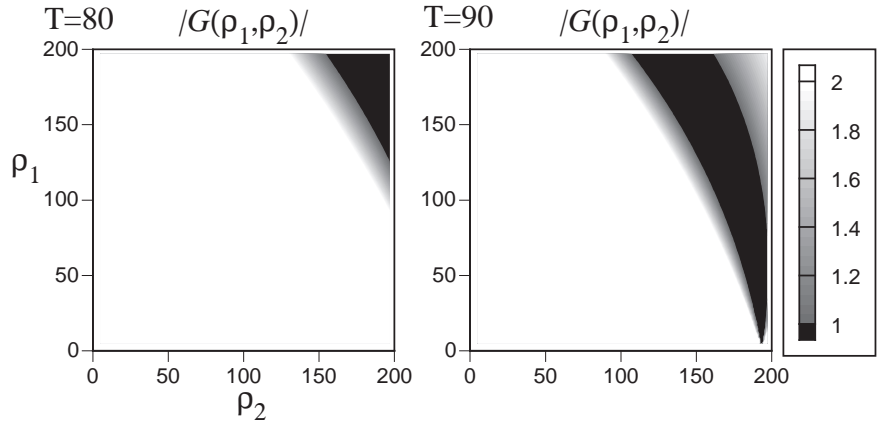}}
 \caption{Contour plots of $\absv{G(\rho_1,\rho_2)}$ at $T=80$ and 
 $90$. $\absv{G(\rho_1,\rho_2)}$ is satisfied in the black region.
 }
 \label{fig:7}
\end{figure}

\begin{figure}
 \epsfxsize=11.6cm
 \centerline{\epsfbox{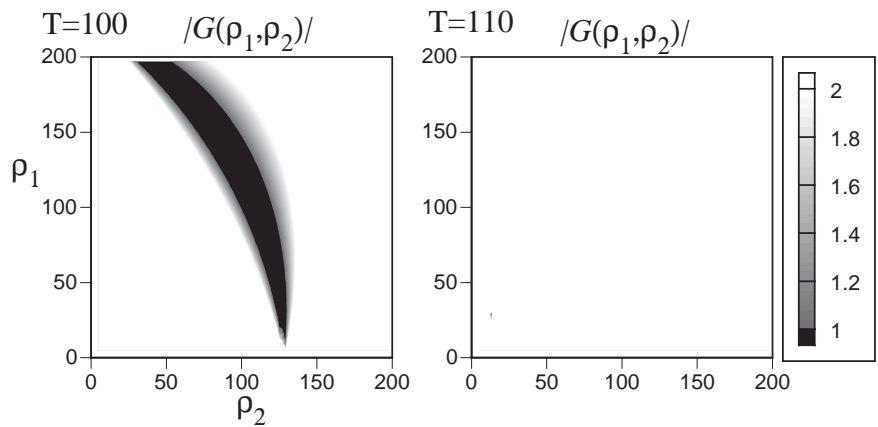}}
 \caption{The same as Fig.~\protect\ref{fig:7} at $T=100$ and $110$. 
 }
 \label{fig:8}
\end{figure}

\noindent
(2) $\tan\beta_0=5$\\
Now $T_{\rm low}=312.48$, so that we ignore the charged Higgs 
contributions.
The behaviors of the effective parameters are qualitatively
the same as the example above, as depicted in Figs.~\ref{fig:5}
and \ref{fig:6}.
Tthe contributions from the charginos and neutralinos are identical
to those above as obvious from (\ref{eq:m32-gaugino}) -- 
(\ref{eq:l6-gaugino}).
The larger $\tan\beta_0$ implies the smaller $y_t$ for a fixed $m_t$,
that is, the smaller $\absv{\Delta_\st m_3^2(0)}$, which implies
the larger $\left(m_3^2\right)_{\rm eff}$ for $\mu A_t>0$.
This lowers the temperature at which $\absv{G(\rho_1,\rho_2)}<1$
is satisfied for some $(\rho_1,\rho_2)$. In this case, we have
\begin{equation}
 B_2/T = 7.368276\times10^{-6},\qquad
 C_1/T = -2.313638\times10^{-3},
\end{equation}
which implies $F(\rho_1,\rho_2)>0$ for $\rho_2>5.104$ at $T=100$,
that is, $F(\rho_1,\rho_2)>0$ holds in the whole region in which
$\absv{G(\rho_1,\rho_2)}<1$ as shown in Figs.~\ref{fig:7} and
\ref{fig:8}.
\par\noindent
{\bf (ii) $\mu A_t<0$}\\
Although the stop contributions change their signs,
those from the charginos and neutralinos are still dominant
for the parameters in Table~\ref{table:2}.
This makes the temperature dependences of all the effective parameters
milder than those in the case of $\mu A_t>0$.
That is, for a wider range of temperature, the conditions for the
spontaneous $CP$ violation will be satisfied.\par
\begin{figure}
 \epsfxsize=11.6cm
 \centerline{\epsfbox{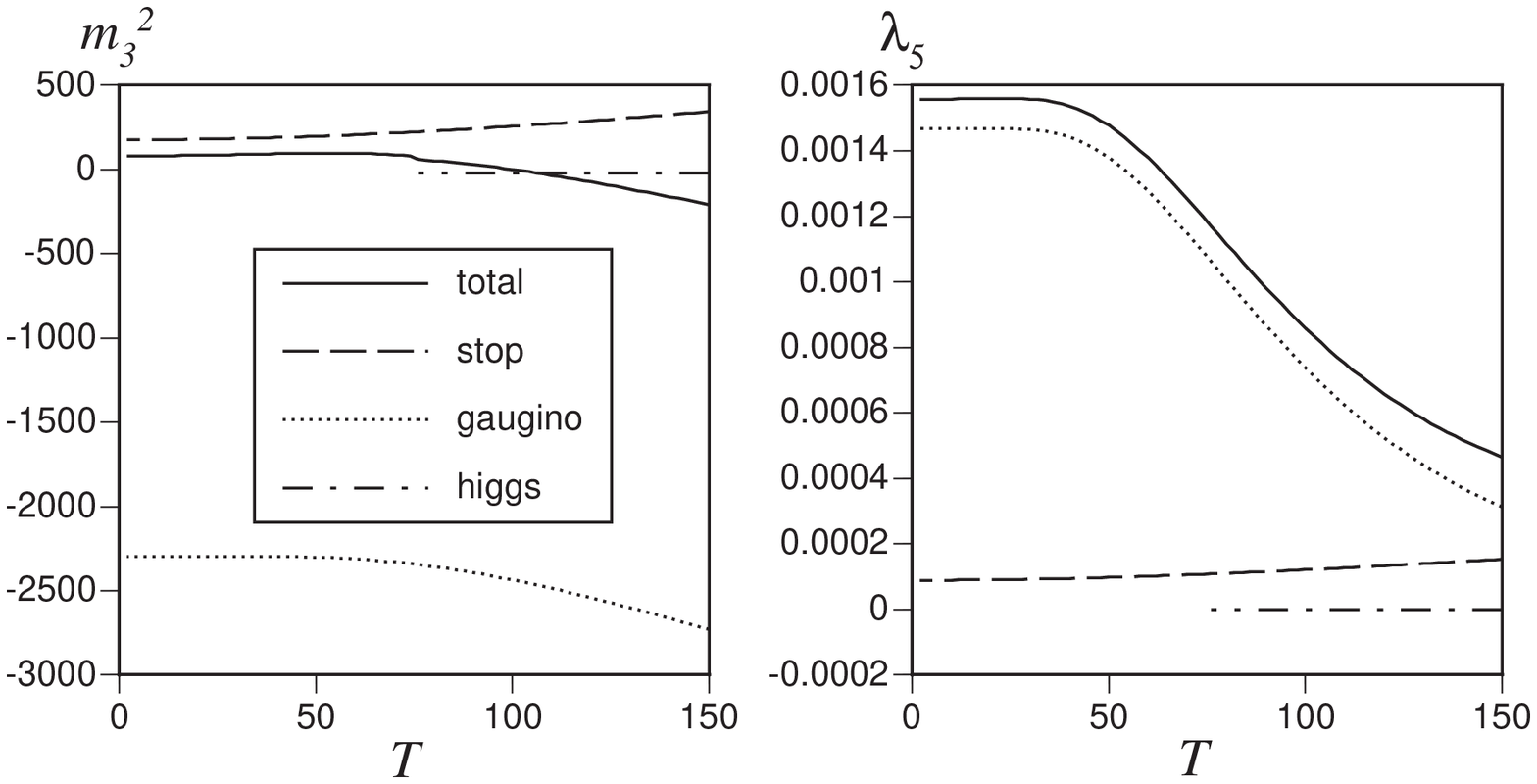}}
 \caption{$\left(m_3^2\right)_{\rm eff}$ and $\lambda_5$ as functions
 of temperature $T$. The total values are given by the solid curves, the 
 corrections from the stop, chargino-neutralino and the charged Higgs 
 bosons are depicted by the dashed, dotted and dotted-dashed curves, 
 respectively. For $T<T_{\rm low}=75.3$ the Higgs contributions are
 ignored.
 }
 \label{fig:9}
\end{figure}
\begin{figure}
 \epsfxsize=11.6cm
 \centerline{\epsfbox{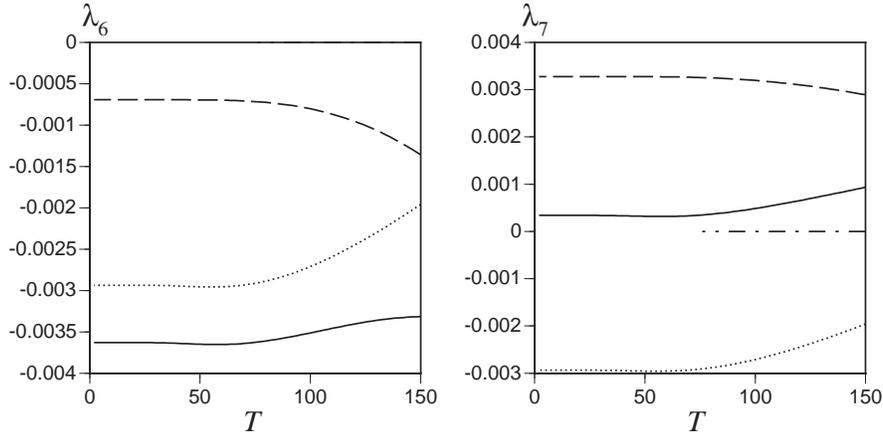}}
 \caption{$\lambda_6$ and $\lambda_7$ as functions of temperature.
 }
 \label{fig:10}
\end{figure}
\begin{figure}
 \epsfxsize=11.6cm
 \centerline{\epsfbox{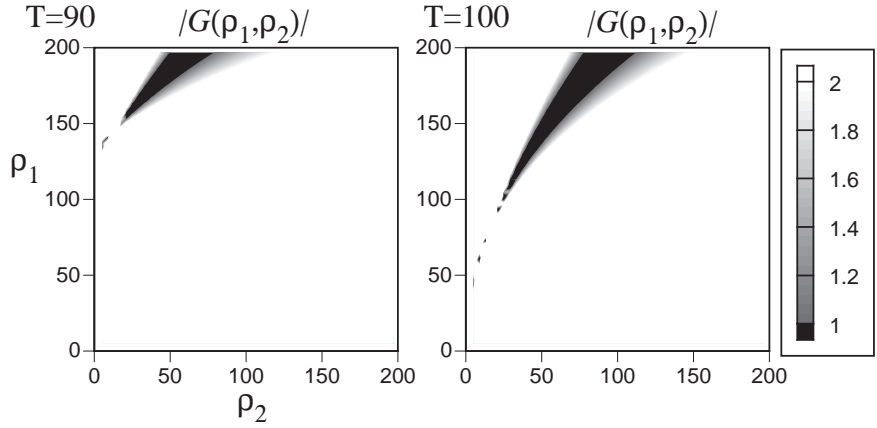}}
 \caption{Contour plots of $\absv{G(\rho_1,\rho_2)}$ at $T=90$ and 
 $100$. $\absv{G(\rho_1,\rho_2)}$ is satisfied in the black region.
 }
 \label{fig:11}
\end{figure}
\begin{figure}
 \epsfxsize=11.6cm
 \centerline{\epsfbox{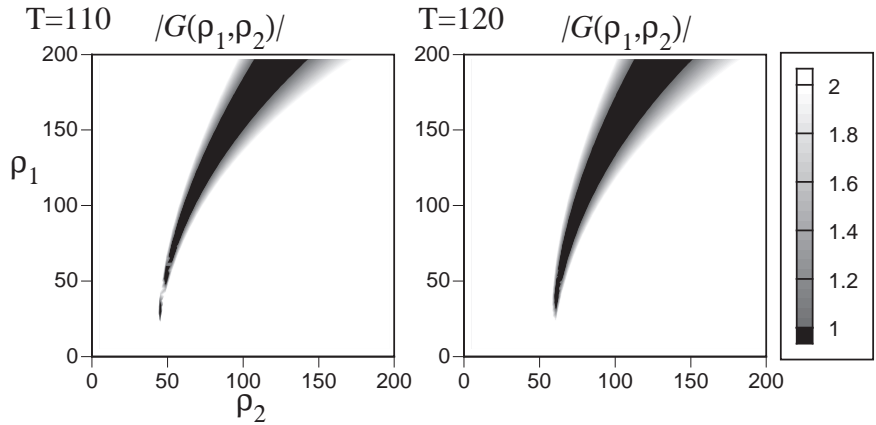}}
 \caption{The same as Fig.~\protect\ref{fig:11} at $T=110$ and $120$. 
 }
 \label{fig:12}
\end{figure}

\begin{table}{l}{h}
 \caption{The parameters used in the numerical analysis in the case
 of $\mu A_t<0$.}
 \label{table:2}
 \begin{center}
  \begin{tabular}{ccccc}
   \hline
    $m_3^2$&$A_t$& $M_2$ & $\mu$ &$m_\sq$ \\
   \hline
    $2200$ & $10$& $300$ &$-300$ &  $400$ \\
   \hline
  \end{tabular}
 \end{center}
\end{table}
\noindent
(1) $\tan\beta_0=1.2$\\
For this, $T_{\rm low}=75.313$, below which the Higgs contributions 
are neglected. The effective parameters are plotted in
Figs.~\ref{fig:9} and \ref{fig:10}.
Since
\begin{equation}
 B_2/T = 3.447779\times10^{-5},\qquad
 C_1/T = 7.269727\times10^{-3},
\end{equation}
$F(\rho_1,\rho_2)>0$ for $\rho_2>16.05$ at $T=100$.
The almost whole region where $\absv{G(\rho_1,\rho_2)}<1$
satisfies this condition as well, as shown in Figs.~\ref{fig:11}
and \ref{fig:12}.
The positive $C_1$ requires smaller $m_3^2$ than that for $C_1<0$
to make $2\left(m_3^2\right)_{\rm eff}+2C_1\rho_2$ nearly equal to
$\lambda_6\rho_1^2+\lambda_7\rho_2^2$.\par
\begin{figure}
 \epsfxsize=11.6cm
 \centerline{\epsfbox{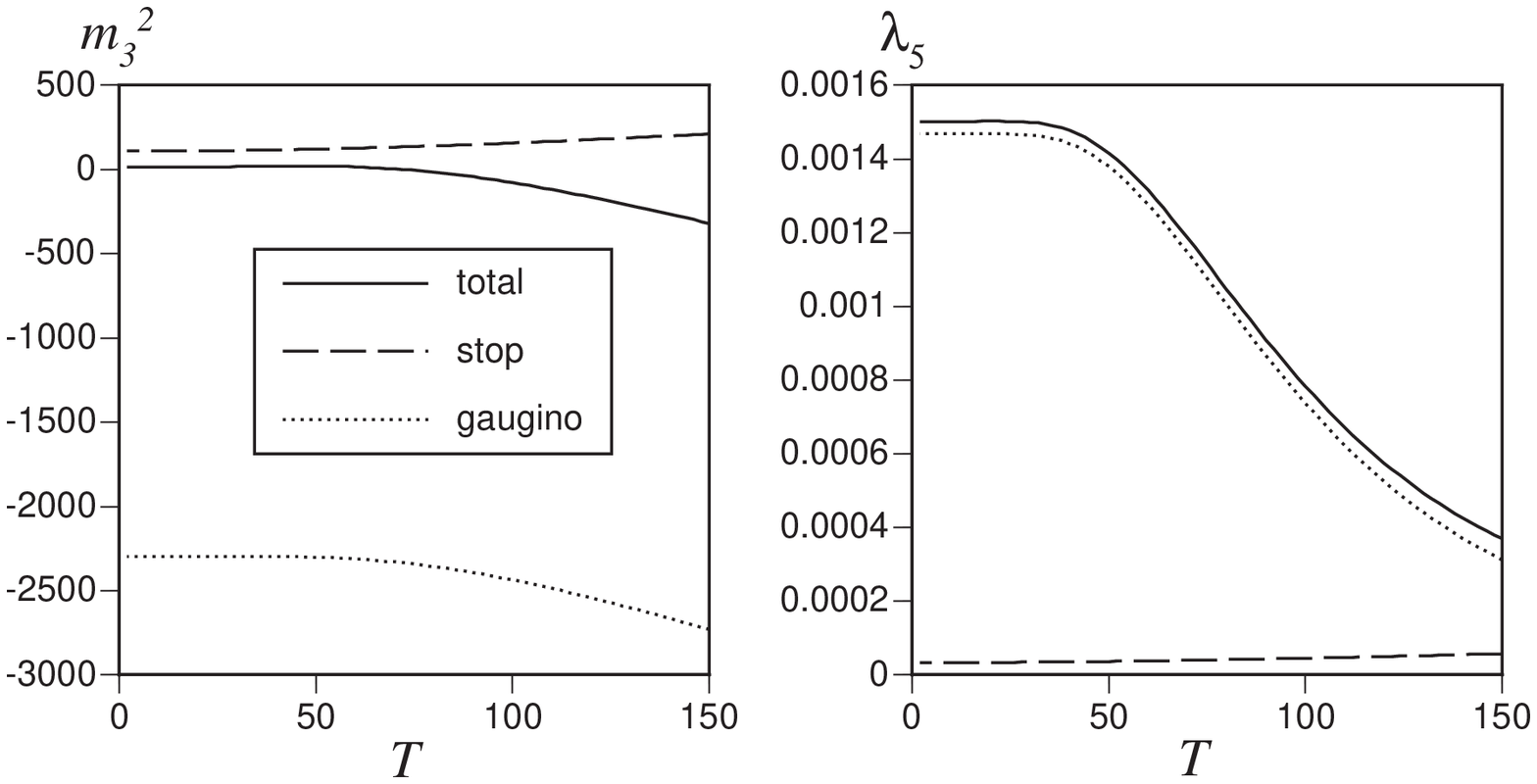}}
 \caption{$\left(m_3^2\right)_{\rm eff}$ and $\lambda_5$ as functions
 of temperature. The total values are given by the solid curves, the 
 corrections from the stop and chargino-neutralino are depicted by
 the dashed and dotted curves, respectively.
 }
 \label{fig:13}
\end{figure}

\begin{figure}
 \epsfxsize=11.6cm
 \centerline{\epsfbox{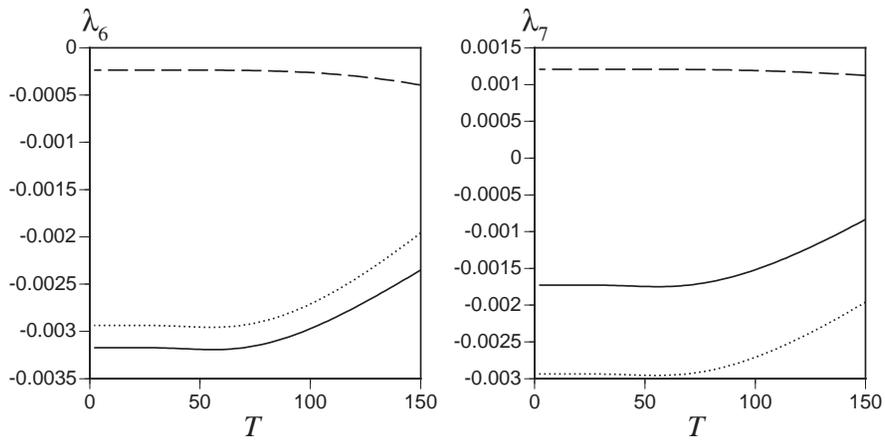}}
 \caption{$\lambda_6$ and $\lambda_7$ as functions
 of temperature.
 }
 \label{fig:14}
\end{figure}

\begin{figure}
 \epsfxsize=11.6cm
 \centerline{\epsfbox{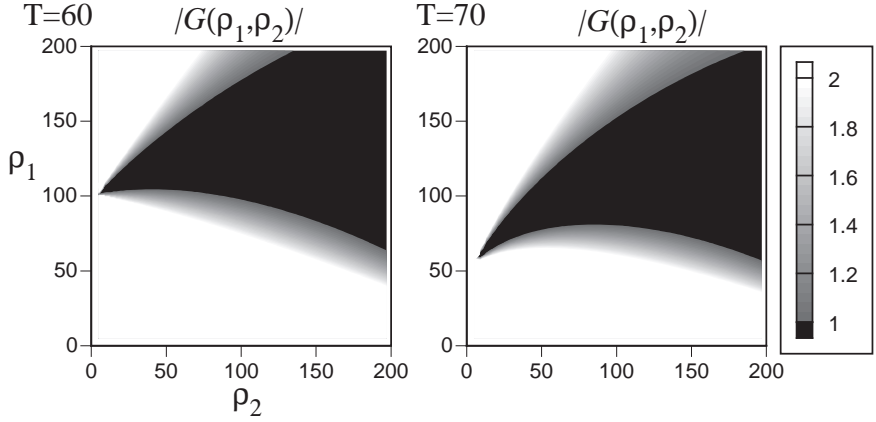}}
 \caption{Contour plots of $\absv{G(\rho_1,\rho_2)}$ at $T=60$ and 
 $70$. $\absv{G(\rho_1,\rho_2)}$ is satisfied in the black region.
 }
 \label{fig:15}
\end{figure}

\begin{figure}
 \epsfxsize=11.6cm
 \centerline{\epsfbox{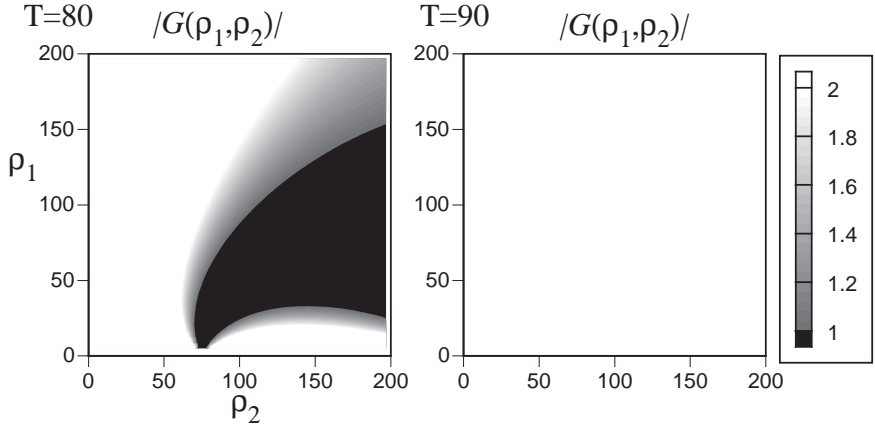}}
 \caption{The same as Fig.~\protect\ref{fig:15} at $T=80$ and $90$. 
 }
 \label{fig:16}
\end{figure}

\noindent
(2) $\tan\beta_0=5$\\
Now $T_{\rm low}=313.08$. We completely ignore the Higgs contributions
in the plots of the effective parameters in Figs.~\ref{fig:13} and
\ref{fig:14}.
By the same reasoning as in the case of $\mu A_t>0$, 
the temperature at which $\absv{G(\rho_1,\rho_2)}<1$ is lowered.
At the same time, $\lambda_5$ becomes larger at lower temperatures
so that the region with $\absv{G(\rho_1,\rho_2)}<1$ grows as seen
from Figs.~\ref{fig:15} and \ref{fig:16}.
For this parameter set, we have
\begin{equation}
 B_2/T = 1.657862\times10^{-5},\qquad
 C_1/T = 3.470457\times10^{-3},
\end{equation}
which implies $F(\rho_1,\rho_2)>0$ for $\rho_2>8.463$ at $T=100$.
\section{Discussions}
We have investigated the possibility of the new type of spontaneous $CP$ 
violation, which occurs at finite temperature in the transient region 
from the symmetric phase to the broken phase separated by the 
electroweak bubble wall.
Since this type of $CP$ violation disappears in the broken phase at
zero temperature, it is free from any constraint on $CP$ violation
from the experiments.
Further it will enhance the generated baryon number by the electroweak 
baryogenesis mechanism. 
Although the $CP$-conjugated pair of the bubbles degenerate in their 
energies as well as their nucleation rates, a tiny explicit $CP$ 
violation consistent with the observation such as the neutron EDM
is sufficient to resolve the degeneracy and to leave the present
BAU\cite{FKOTa}.\par
For this mechanism to work, some constraints are imposed on the 
parameters in the MSSM. First of all, $\mu M_2<0$ is required
to make $\left(m_3^2\right)_{\rm eff}$ decrease by the chargino and
neutralino contributions as noted in the previous section.
$\absv{\mu}\sim\absv{M_2}$ yields a positive correction to $\lambda_5$
by these particles up to $T\simeq 100\mbox{GeV}$.
For the expansion used for the light stop contributions to be valid,
$\mu^2<m_\sq^2$ and $A_t^2<m_\sq^2$ are required.
Together with $\absv{\mu}\sim\absv{M_2}$, these conditions make the
stop contributions smaller than those from the charginos and neutralinos.
The $\theta$-dependent $\rho^3$-terms are induced only if one of the 
soft-supersymmetry-breaking masses of the stop almost vanishes; 
$m_\st\simeq 0$.
Among the coefficients of these terms, $C_1$ gives contributions
to the numerator of $G(\rho_1,\rho_2)$ comparable to those from the
charginos and neutralinos.
Whether $\absv{G(\rho_1,\rho_2)}<1$ is satisfied is sensitive to
$\left(m_3^2\right)_{\rm eff}$ in its numerator.
Since $2C_1\rho_2$ appears in the numerator with the same sign as
$\left(m_3^2\right)_{\rm eff}$, the effect of positive $C_1$ ($\mu A_t<0$)
is compensated by reducing the tree-level $m_3^2$ as shown in the 
two examples in the previous section.\par
One might wonder how the difficulty encountered in the case of
the spontaneous $CP$ violation at zero temperature\cite{Maekawa}\  
is avoided.
$\lambda_5>0$ is satisfied at $T=0$ if the chargino and neutralino 
contributions dominate over those from the Higgs bosons and stops.
This also applies to $F(\rho_1,\rho_2)$ in our case, except for
the Higgs boson contribution, which is small since the effective
$m_1^2m_2^2$ is larger than $m_3^4$ at $T\simeq T_C$.
The problem was that for $\absv{G(\rho_1,\rho_2)}<1$ to be satisfied at
$T=0$, where $B_{1,2}=C_{1,2}=0$, $\left(m_3^2\right)_{\rm eff}$
must be so small as 
$(\lambda_6\cos^2\beta+\lambda_7\sin^2\beta)v_0^2$, which implies
that the pseudoscalar boson is too light.
Its mass $m_A$ is related to $m_3^2$ at the tree level by
$m_A^2=(\tan\beta_0+\cot\beta_0)m_3^2$. The smallest value of this 
tree-level $m_A$ in our examples is $m_A=67\mbox{GeV}$ for
$m_3^2=2200\mbox{GeV}^2$ and $\tan\beta_0=1.2$.
However, $m_A$ should be calculated at the minimum of the corrected
potential. The value of $m_A$ calculated in this way might be
sufficiently large for suitable range of the parameters.
Further the parameters of the theory are not so constrained in our
case compared to the case at $T=0$. This is because, 
if only the conditions are satisfied at some $(\rho_1,\rho_2)$ in the 
transient region, the $CP$ phase could be large enough to generate 
the BAU around the bubble wall.\par
Finally we emphasize that the spontaneously-$CP$-breaking minimum
does not have to be the global minimum of $V_{\rm eff}$.
The transitional $CP$ violation could take place if the conditions
are satisfied for some fixed $(\rho_1,\rho_2)$, since such a bubble 
wall with transitional $CP$ violation would have a lower energy than
that without $CP$ violation.
Hence we do not need to be afraid that such a local minimum may not be 
the absolute minimum. For another reason, however, we should
understand the global structure of the effective potential, which
determines $T_C$, to know whether the transitional $CP$ violation occurs
or not.
In this sense, the conditions we examined here should be regarded
as the necessary conditions but not the sufficient ones.
With the knowledge of the global structure of $V_{\rm eff}$, one could
find the $CP$-violating profile of the bubble wall so that one could
estimate the generated baryon number.
\section*{Acknowledgments}
This work is supported in part by Grant-in-Aid for Scientific
Research on Priority Areas (Physics of $CP$ violation, No.09246223),
No.09740207 (K.F.) 
and No.09640378 (F.T.) from the Ministry of Education, Science, 
and Culture of Japan.
\appendix
\section{Loop Corrections}
Here we summarize the expressions  for the contributions to the effective
parameters in terms of the finite-temperature Feynman integrals from
the charginos, neutralinos, charged Higgs bosons and stops
at finite temperature.
We also show various formulas to calculate the Feynman integrals.
According to the definitions of the effective parameters
(\ref{eq:def-eff-m32}) -- (\ref{eq:def-eff-l7}), the contribution of
each particle is expressed in terms of the propagators in the 
symmetric phase ($\rho_i=0$) and the vertices which are related to 
the derivatives of the mass matrices.\par
The contributions from the charginos whose mass matrix is given by
(\ref{eq:chargino-mass-matrix}) to the effective parameters are
\begin{eqnarray}
 \Delta_{\chi^\pm} m_3^2 &=&
 2 g_2^2 \mu M_2\,i\int_k^{(+)}\Delta_1(k)\Delta_2(k), \\
 \Delta_{\chi^\pm} \lambda_5 &=&
 -2 g_2^4 (\mu M_2)^2\,i\int_k^{(+)}\Delta_1^2(k)\Delta_2^2(k), \\
 \Delta_{\chi^\pm} \lambda_6 &=& \Delta_{\chi^\pm} \lambda_7 =
 2 g_2^4 \mu M_2\,i\int_k^{(+)} k^2\Delta_1^2(k)\Delta_2^2(k),
\end{eqnarray}
where
\begin{equation}
 \Delta_1(k) = {1\over{k^2-M_2^2}},\qquad
 \Delta_2(k) = {1\over{k^2-\mu^2}},
\end{equation}
and the integral implies
\begin{equation}
 \int_k^{(\mp)} \equiv
 iT\sum_{n=-\infty}^\infty \int{{d^3\mbox{\bf k}}\over{(2\pi)^3}}
 \quad\mbox{with }
 \left\{\begin{array}{ll}
   k_0=i\omega_n = 2ni\pi/T \quad &  \mbox{for bosons} \\
   k_0=i\omega_n = (2n+1)i\pi/T\quad &\mbox{for fermions}.
 \end{array}\right.
\end{equation}
\par
The neutralino contributions are rather lengthy because of its
mass matrix given by (\ref{eq:neutralino-mass-matrix}):
\begin{eqnarray}
 \Delta_{\chi^0} m_3^2 &=&
 2(g_2^2+g_1^2) \mu\,i\int_k^{(+)}\Delta_1(k)\Delta_2(k)\Delta_3(k)
  \left[(k^2-P^2)Q^2 + PR^2\right], \\
 \Delta_{\chi^0} \lambda_5 &=&
 -4(g_2^2+g_1^2)^2 \mu^2\,i\int_k^{(+)}
  \Delta_1^2(k)\Delta_2^2(k)\Delta_3^2(k)(k^2-P^2-R^2)  \nonumber\\
  & &{}
  \times\left[(k^2-P^2)Q^2 + PQR^2 \right], \\
 \Delta_{\chi^0} \lambda_6 &=& \Delta_{\chi^0} \lambda_7 =
 2(g_2^2+g_1^2) \mu\,i\int_k^{(+)}
  \Delta_1^2(k)\Delta_2^2(k)\Delta_3^2(k) k^2(k^2-P^2-R^2) \nonumber\\
  & &{}
  \times\left[(2k^2-2P^2-R^2)Q + PR^2 \right],
\end{eqnarray}
where
\begin{eqnarray}
 \Delta_1(k) &=& {1\over{k^2-\mu_1^2}},\quad
 \Delta_2(k) = {1\over{k^2-\mu_2^2}},\quad
 \Delta_3(k) = {1\over{k^2-\mu^2}},   \\
 \mu^2_{1,2} &=&
 {{P^2+Q^2+2R^2\pm\absv{P+Q}\sqrt{(P-Q)^2+4R^2}}\over2},
\end{eqnarray}
with
\begin{eqnarray}
 P &=& M_2\sin^2\theta_W + M_1\cos^2\theta_W, \\
 Q &=& M_2\cos^2\theta_W + M_1\sin^2\theta_W, \\
 R &=& (M_2-M_1)\sin\theta_W\cos\theta_W.
\end{eqnarray}
If $M_1=M_2$, $P=Q=M_2$ and $R=0$ so that $\mu_1=\mu_2=M_2$.
In this case the neutralino contributions are reduced to
\begin{eqnarray}
 \Delta_{\chi^0} m_3^2 &=&
 {1\over{\cos^2\theta_W}} \Delta_{\chi^\pm} m_3^2,   \\
 \Delta_{\chi^0} \lambda_5 &=&
 {2\over{\cos^4\theta_W}} \Delta_{\chi^\pm} \lambda_5, \\
 \Delta_{\chi^0} \lambda_6 &=& 
 \Delta_{\chi^0} \lambda_7 =
 {2\over{\cos^4\theta_W}} \Delta_{\chi^\pm} \lambda_6 =
 {2\over{\cos^4\theta_W}} \Delta_{\chi^\pm} \lambda_7.
\end{eqnarray}
We consider this special case of $M_2=M_1$ for simplicity.\par
The corrections to the effective parameters from the charged Higgs
bosons are
\begin{eqnarray}
 \Delta_{\phi^\pm}m_3^2 &=&
 -\half g_2^2 m_3^2\,i\int_k^{(-)}\Delta_1(k)\Delta_2(k),   \\
 \Delta_{\phi^\pm}\lambda_5 &=&
 {1\over4}g_2^4(m_3^2)^2\,i\int_k^{(-)}\Delta_1^2(k)\Delta_2^2(k),\\
 \Delta_{\phi^\pm}\lambda_6 &=&
 -{1\over4}g_2^4 m_3^2\,i\int_k^{(-)}\Delta_1^2(k)\Delta_2^2(k)
  \nonumber\\
 & &{}\times
 \left[k^2-{{m_1^2}\over{2\cos^2\theta_W}}-
  \left(1-{1\over{2\cos^2\theta_W}}\right)m_2^2 \right],   \\
 \Delta_{\phi^\pm}\lambda_7 &=&
 -{1\over4}g_2^4 m_3^2\,i\int_k^{(-)}\Delta_1^2(k)\Delta_2^2(k)
  \nonumber\\
 & &{}\times
 \left[k^2-\left(1-{1\over{2\cos^2\theta_W}}\right)m_1^2
 -{{m_2^2}\over{2\cos^2\theta_W}} \right],
\end{eqnarray}
where
\begin{eqnarray}
 \Delta_i(k) &=& {1\over{k^2-\mu_i^2}},\\
 \mu_{1,2}^2 &=& {{m_1^2+m_2^2\pm\sqrt{(m_1^2-m_2^2)^2+4m_3^4}}\over2}.
\end{eqnarray}
In practice, we use the daisy-corrected ${\bar m}_{1,2}^2$ defined by
(\ref{eq:def-m-bar}) instead of $m_{1,2}^2$.\par
For the stop with $m_\st=0$, we should treat the finite-temperature 
corrections from the heavy and light mass eigenstates separately.
When $m_\st\not=0$, the stop contributions are given by the integrals below.
\begin{eqnarray}
 \Delta_\st m_3^2 &=&
 -N_c y_t^2 \mu A_t\,i\int_k^{(-)}\Delta_1(k)\Delta_2(k), \\
 \Delta_\st \lambda_5 &=&
 N_c y_t^4 (\mu A_t)^2\,i\int_k^{(-)}\Delta_1^2(k)\Delta_2^2(k),\\
 \Delta_\st \lambda_6 &=&
 -N_c y_t^2 \mu A_t\,i\int_k^{(-)}
 \Bigl[ -{1\over4}\left({{g_1^2}\over3}-g_2^2\right)\Delta_1^2(k)\Delta_2(k)
 \nonumber\\
 & &{}+{{g_1^2}\over3}\Delta_1(k)\Delta_2^2(k)
 +y_t^2\mu^2\Delta_1^2(k)\Delta_2^2(k) \Bigr], \\
 \Delta_\st \lambda_7 &=&
 -N_c y_t^2 \mu A_t\,i\int_k^{(-)}
 \Bigl[\left({1\over4}\left({{g_1^2}\over3}-g_2^2\right)+y_t^2\right)
 \Delta_1^2(k)\Delta_2(k)\nonumber\\
 & &{}+\left(-{{g_1^2}\over3}+y_t^2\right)\Delta_1(k)\Delta_2^2(k)
 +y_t^2A_t^2\Delta_1^2(k)\Delta_2^2(k) \Bigr],
\end{eqnarray}
where
\begin{equation}
 \Delta_1(k) = {1\over{k^2-m_\sq^2}},\qquad
 \Delta_2(k) = {1\over{k^2-m_\st^2}}.
\end{equation}
The zero-temperature corrections can also be extracted from these
integrals by use of the formulas given below.
For $m_\st^2=0$, we need the infrared cutoff to regularize the 
zero-temperature integrals to obtain the results (\ref{eq:m32-stop2})
-- (\ref{eq:l7-stop2}).
The finite-temperature Feynman integrals can be divided into
the zero-temperature ones and the finite-temperature correction to them,
which are usually expressed in terms of one-dimensional integrals.
The integral appearing in the corrections to $m_3^2$ is simplified
by the formula
\begin{eqnarray}
 -i\int_k^{(\mp)}\Delta_1(k)\Delta_2(k) &=&
 \int{{d^4k_E}\over{(2\pi)^4}}{1\over{(k^2+m_1^2)(k^2+m_2^2)} }\pm
 {1\over{2\pi^2}}f_2^{(\mp)}\!\left({m_1\over T},{m_2\over T}\right)
\nonumber\\
 &=&L(m_1,m_2)
\pm{1\over{2\pi^2}}f_2^{(\mp)}\!\left({m_1\over T},{m_2\over T}\right),
\end{eqnarray}
where the zero-temperature part is explicitly evaluated to yield
\begin{equation}
 L(m_1,m_2)= {1\over{16\pi^2}}\left[ 1- 
   {m_1^2\over{m_1^2-m_2^2}}\log{m_1^2\over{M_{\rm ren}^2}}+
   {m_2^2\over{m_1^2-m_2^2}}\log{m_2^2\over{M_{\rm ren}^2}}\right]>0,
                 \label{eq:def-L}
\end{equation}
which is renormalized by the $\overline{\rm DR}$-scheme just as
(\ref{eq:V1-zero-T}). $f_2^{(\mp)}(a,b)$ in the finite-temperature part 
is given by
\begin{eqnarray}
 f_2^{(\mp)}(a,b) &=&-{1\over{a^2-b^2}}\int_0^\infty dx\Bigl(
  {{x^2}\over\sqrt{x^2+a^2}}{1\over{e^{\sqrt{x^2+a^2}}\mp1}} 
  -{{x^2}\over\sqrt{x^2+b^2}}{1\over{e^{\sqrt{x^2+b^2}}\mp1}}\Bigr),
                  \nonumber\\
 & &              \label{eq:def-f2-1}
\end{eqnarray}
for $a\not=b$ and
\begin{equation}
 f_2^{(\mp)}(a,a)=\half\int_0^\infty{1\over{\sqrt{x^2+a^2}}}
                  {1\over{e^{\sqrt{x^2+a^2}}\mp1}}.
                        \label{eq:def-f2-2}
\end{equation}
$f_2^{(\mp)}(a,b)$ is positive for any $(a,b)$.\par
The integrals in the corrections to $\lambda_{6,7}$ are calculated by 
use of the following integral: 
\begin{equation}
 -i\int_k^{(\mp)}\Delta_1^2(k)\Delta_2(k) =
 {1\over{16\pi^2m_1^2}}H\left({{m_1^2}\over{m_2^2}}\right)\pm
 {1\over{2\pi^2T^2}}f_3^{(\mp)}\!\left({m_1\over T},{m_2\over T}\right),
\end{equation}
where
\begin{equation}
 H(\alpha) = {\alpha\over{\alpha-1}}\left(
    {1\over{\alpha-1}}\log\alpha -1 \right),   \label{eq:def-H}
\end{equation}
with $H(1)=-1/2$ and
\begin{equation}
 f_3^{(\mp)}(a,b) =
 {1\over{2(a^2-b^2)}}\int_0^\infty{{dx}\over{\sqrt{x^2+a^2}}}
    {1\over{e^{\sqrt{x^2+a^2}}\mp1}} -
 {1\over{a^2-b^2}}f_2^{(\mp)}(a,b),   \label{eq:def-f3-1}
\end{equation}
for $a\not=b$ and
\begin{eqnarray}
 f_3^{(\mp)}(a,a) &=&
 -{1\over8}\int_0^\infty dx\Biggl[
 {1\over{(x^2+a^2)^{3/2}}}{1\over{e^{\sqrt{x^2+a^2}}\mp1}} +
 {1\over{x^2+a^2}}{{e^{\sqrt{x^2+a^2}}}\over{(e^{\sqrt{x^2+a^2}}\mp1)^2}}
  \Biggr].          \label{eq:def-f3-2}
\end{eqnarray}
$f_3^{(\mp)}(a,b)$ is negative for any $(a,b)$.\par
The integral in the corrections to $\lambda_5$ is reduced to
\begin{equation}
 -i\int_k^{(\mp)}\Delta_1^2(k)\Delta_2^2(k) =
 {1\over{16\pi^2m_1^4}}K\left({{m_1^2}\over{m_2^2}}\right)\pm
 {1\over{2\pi^2T^4}}f_4^{(\mp)}\!\left({m_1\over T},{m_2\over T}\right),
\end{equation}
where
\begin{equation}
 K(\alpha) = {\alpha\over{(\alpha-1)^2}}\left(
  {{\alpha+1}\over{\alpha-1}}\log\alpha -2 \right),   \label{eq:def-K}
\end{equation}
with $K(1)=1/6$ and
\begin{eqnarray}
 f_4^{(\mp)}(a,b)&=&
 {1\over{2(a^2-b^2)^2}}\int_0^\infty dx\Bigl(
   {1\over{\sqrt{x^2+a^2}}}{1\over{e^{\sqrt{x^2+a^2}}\mp1}} 
   \nonumber\\
 & &\qquad +
   {1\over{\sqrt{x^2+b^2}}}{1\over{e^{\sqrt{x^2+b^2}}\mp1}} \Bigr)
   -{2\over{(a^2-b^2)^2}} f_2^{(\mp)}(a,b), \label{eq:def-f4-1}
\end{eqnarray}
for $a\not=b$ and
\begin{eqnarray}
 f_4^{(\mp)}(a,a) &=&
 {1\over{16}}\int_0^\infty dx\Biggl\{
 {1\over{(x^2+a^2)^{5/2}}}{1\over{e^{\sqrt{x^2+a^2}}\mp1}} +
 \Bigl[{1\over{(x^2+a^2)^2}}      \nonumber\\
  & &\qquad\qquad
  -{1\over{3(x^2+a^2)^{3/2}}}\Bigr]
   {{e^{\sqrt{x^2+a^2}}}\over{(e^{\sqrt{x^2+a^2}}\mp1)^2}}\nonumber\\
 & &\qquad\qquad
  +{2\over{3(x^2+a^2)^{3/2}}}
  {{e^{2\sqrt{x^2+a^2}}}\over{(e^{\sqrt{x^2+a^2}}\mp1)^3}}\Biggr\}.
                       \label{eq:def-f4-2}
\end{eqnarray}
$f_4^{(\mp)}(a,b)$ is positive for any $(a,b)$.\par
%
%
%

\end{document}